\documentclass[twocolumn,amssymb,amsmath,nofootinbib,aps,prd]{revtex4-2}

\usepackage{aas_macros}
\usepackage{physics}

\usepackage{graphicx}% Include figure files
\usepackage{dcolumn}% Align table columns on decimal point
\usepackage{bm}% bold math

\usepackage[dvipsnames,table]{xcolor}
\usepackage{rotating}

\definecolor{LightCyan}{rgb}{0.88,1,1}

\usepackage{hyperref}

\hypersetup{
  colorlinks   = true, %Colours links instead of ugly boxes
  urlcolor     = ForestGreen, %Colour for external hyperlinks
  linkcolor    = OrangeRed, %Colour of internal links
  citecolor    = blue %Colour of citations
}

\begin{document}

\title{Anisotropic universe with anisotropic dark energy}% Force line breaks with \\

\author{Anshul Verma$^1$}
\email{anshulverma.rs.phy19@iitbhu.ac.in}

\author{Pavan K. Aluri$^1$}%
\email{pavanaluri.phy@iitbhu.ac.in}
\affiliation{$^1$Department of Physics, Indian Institute of Technology (BHU), Varanasi - 221005, India.}

\author{David F. Mota$^2$}
\email{d.f.mota@astro.uio.no}
\affiliation{$^2$Institute of Theoretical Astrophysics, University of Oslo, P.O. Box 1029 Blindern, N-0315 Oslo, Norway.}

\date{\today}% It is always \today, today,

\begin{abstract}
We investigate the anisotropic parameterization of the dark energy equation of state within the framework of an axisymmetric (planar) Bianchi-I universe. Using the latest Pantheon+ Type Ia Supernova dataset, augmented by SH0ES Cepheid distance calibrators, we constrain both the equation of state for anisotropic dark energy and other standard cosmological parameters. Additionally, we examine the presence of an underlying anisotropic axis. Our analysis yields a mean anisotropic dark energy equation of state of  $\bar{w} = -0.86^{+0.15}_{-0.11}$ and a difference in the equation of states in and perpendicular to the plane of the axisymmetric Bianchi-I spacetime of $\delta_w = -0.129^{+0.090}_{-0.064}$. We also identify an axis of anisotropy at approximately $(272^{\circ}, 21^{\circ})$ in galactic coordinates. Through a comparative study of different cosmological models, we find that the data favor a Bianchi-I universe with anisotropic dark energy, where the equation of state deviates from ``-1'' along the axis of anisotropy (the $w_b$CDM model), over both other anisotropic models considered and the standard flat $\Lambda$CDM or $w$CDM models.
\end{abstract}

\maketitle

%\tableofcontents

\section{Introduction}
\label{sec:level1}
Understanding the nature and behaviour of dark energy stands as one of the most pressing challenges in modern cosmology. While the standard cosmological model, based on the Friedmann-Lema\^itre-Robertson-Walker (FLRW) metric, has been remarkably successful in explaining a wide range of observational data, it assumes spatial homogeneity and isotropy on large scales of our universe also known as the Cosmological principle~\cite{WeinbergGRCosmoBook}.

However, recent observations indicate a violation of Cosmological principle~\cite{schwarz2016,bull2016,perivolaropoulos2022,abdalla2022,aluri2023cp}. Although
observations of cosmic microwave background (CMB) radiation~\cite{cobe1996maps,wmap9yrmaps,plk2018maps} broadly conforms with standard cosmological model based on Cosmological principle,CMB sky exhibited various anisotropic features at large angular scales which were extensively studied in the literature~\cite{wmap7yranom,plk2013isostat,plk2015isostat,plk2018isostat}. Similar tests of statistical (an)isotropy were also conducted on various other astronomical data sets such as Type Ia Supernovae~\cite{schwarz2007sn1a,perivolaropoulos2010sn1a,subir2011sn1a,Kalus:2013,wiltshire2013sn1a,2014wang,appleby2015sn1a,Soltis:2019,zhao2019sn1a}, Quasars~\cite{hutsemekers1998,hutsemekers2001,jain2004,hutsemekers2014,Secrest2021}, radio polarization vector alignments~\cite{birch1982,jain1999,tiwari2013radioalgn}, large scale velocity flows in $\Lambda$CDM cosmology~\cite{watkins2009}, large-scale velocity field from the Cosmicflows-4 data~\cite{Hoffman2024MNRAS}. Interestingly many of these anisotropy axes align with each other broadly in the same direction as the Virgo cluster~\cite{johnjain2004}. In the current era of explosive amounts of data and computation being available readily, the Cosmological principle specifically ``statistical isotropy'' has been put to test from time to time for its validity.

In the standard cosmological model viz., the flat $\Lambda$CDM model, dark energy which is responsible for the current accelerated expansion of our universe~\cite{Perlmutter:1999} is modeled as a cosmological constant, $\Lambda$, that is an isotropic source with an equation of state (EOS) given by $w_{\rm de}=-1$~\cite{Spergel2003ApJS,plk2018cosmopar}.  In light of signals of anisotropy seen in late time probes of our universe (as well as from early universe viz., CMB), it would be interesting to test whether the current expansion of our universe is anisotropic i.e., whether `cosmological constant' (or another form of dark energy) is driving this expansion equally in all directions or not.

In the late universe, anisotropic dark energy (ADE) may provide a natural mechanism to reintroduce anisotropy in light of various instances of isotropy violation cited above.
ADE models which allows for the EOS parameter, $w_{\rm de}$, to be different from ``-1'' or some constant value different from ``-1'' (eg. $w$CDM) in different directions offer a promising avenue for exploring deviations from isotropy and addressing cosmological tensions. Many models were proposed in the literature that induces anisotropy (at early or late times resulting in a rapid expansion phase) such as vector dark energy models, phenomenological parameterization of dark energy, modified gravity, etc., for example~\cite{Armendariz-Picon2004,Bohmer2007,Rodrigues2008,Koivisto2008a,Koivisto2009,Appleby2010,
Zuntz2010,Campanelli2011ade,Thorsrud2012,Akarsu2014,Saha2014,Lavinia2016,Beltran2019,
Motoa-Manzano2021,Orjuela-Quintan2021}.
In this context, the Bianchi-I universe serves as an immediate generalization for studying spatially homogeneous but anisotropic cosmologies~\cite{Jaffe2005ApJ,Pontzen2009,coles2011,saadeh2016MNRAS,koivisto2008,2011campanelli,2013appleby,Wang:2018,Tedesco2018,Akarsu:2019,Amirhashchi:2020,2021yadav,2022rahman,2023Akarsu}.

To begin with, we are looking for a space-time metric $g_{\mu\nu}$ which can describe global anisotropy that is a solution to Einstein's field equations but is homogeneous. There exists such family of solutions called ``Bianchi models'' or ``homogeneous cosmologies''~\cite{maccallumellis1969,maccallumellis1970}. In fact, there has been extensive work done on these anisotropic but homogeneous models~\cite{RyanShepley1975,Wainwright1997,Stephani2003,Coley2003}.
An axisymmetric Bianchi-I metric provides sufficient features for investigating the interplay between geometry and energy content, offering insights into the underlying physics driving cosmic evolution. So, a simplest generalization to standard cosmological model based on FLRW metric is to consider a Bianchi type-I metric that is given by
\begin{equation}
ds^2 = c^2 dt^2 - a(t)^2 [dx^2 + dy^2] - b(t)^2 dz^2\,,
\label{eq:b1metric-plnr}
\end{equation}
where the expansion factors are different in the $x$-$y$ plane ($a(t)$) and the $z$-axis normal to it ($b(t)$). The constant `$c$' in the above metric is the speed of light and is set to unity i.e., $c=1$ for the rest of this paper. They could otherwise be different in different directions, say, $a(t)$, $b(t)$ and $c(t)$ along the three Cartesian coordinates. The form of the Bianchi-I metric we chose above is motivated by different kinds of physical sources which exhibit azimuthal symmetry~\cite{barrow1997,barera2004,campanelli2009}. Such a space-time has come to be known as Eccentric or Ellipsoidal Universe \cite{barera2004,campanelli2006}.

The energy-momentum tensor of an anisotropic source with axial symmetry corresponding to the form of the metric in Eq.~(\ref{eq:b1metric-plnr}) is taken to be
\begin{equation}
    T^{\mu}_{\nu} = {\rm diag}(\rho,-p_a,-p_a,-p_b)\,,
\label{eq:b1emtensor}
\end{equation}
where, $p_a$ and $p_b$ are the anisotropic pressure terms in the plane normal to and along the anisotropy axis, and $\rho$ is its energy density. We will consider anisotropic dark energy with different equations of state as a source of anisotropy in this paper, for which $w_a=p_a/\rho_{\rm ade}$ and $w_b=p_b/\rho_{\rm ade}$. For an isotropic source like usual (dark+visible) matter ($m$) and radiation ($r$), $w_a=w_b=w_i=p_i/\rho_i$ where $i=m,r$. As usual $w_m=0$ for dust-like matter and $w_r=1/3$ for radiation. Therefore the total energy-momentum tensor is given by their sum viz., $\left(T^{\mu}_{\nu}\right)_{\rm ade} + \left(T^{\mu}_{\nu}\right)_{m} + \left(T^{\mu}_{\nu}\right)_{r}$. We do not consider radiation hereafter in deriving the model constraints (except briefly while discussing the results).

In section~\ref{sec:efe-Dl}, we briefly describe the Einstein's field equations for the chosen Bianchi-I metric, the relationship between distance modulus and redshift within the context of a Bianchi-I space-time, and finally a system of evolution equations (in dimensionless variables) essential for deriving constraints later in our investigation. Following this, we provide a concise overview of the data set utilized in our analysis in section~\ref{sec:SNIa-data-method} before presenting our findings and a detailed analysis of constraints obtained in section~\ref{sec:results}. Lastly, in section~\ref{sec:concl}, we discuss the results and conclude.

\section{Luminosity distance-vs-Redshift relation in a Bianchi-I universe}
\label{sec:efe-Dl}

To constrain the parameters characterizing Bianchi-I model of Eq.~(\ref{eq:b1metric-plnr}), we utilize the latest compilation of Type Ia Supernovae (SNIa) data as elaborated in the next section. Here, we introduce the theoretical luminosity distance ($d_L$) versus redshift ($z$) relation, which is central to deriving constraints on various cosmological parameters of our model using SNIa data. Additionally, we outline the evolution equations governing these parameters, which must be solved in conjunction with the $d_L-z$ relation. For more details, the reader may consult the Refs.~\cite{koivisto2008,2011campanelli,aluri2013sn1a,anshul2024}.

We start by defining a mean scale factor and eccentricity parameter (denoting deviation from isotropic expansion) for our Bianchi-I model as,
\begin{equation}
A = (a^2 b)^{1/3} \quad \text{and} \quad e^2 = 1 - \frac{b^2}{a^2}\,.
\label{eq:mean-scalefac-ecc}
\end{equation}
These definitions in turn allow us to define average Hubble parameter, and the shear parameter i.e., the differential expansion rate of our universe as,
\begin{equation}
H= (2H_a + H_b)/3 \quad {\rm and} \quad \Sigma = (H_a - H_b)/\sqrt{3}\,,
\label{eq:mean-hubble-shear}
\end{equation}
where $H=\dot{A}/A$, $H_a = \dot{a}/a$ and $H_b=\dot{b}/b$. Here an overdot ($\dot{\hspace{1em}}$) represents derivative with respect to cosmic time $t$.
In terms of these variables, Einstein's field equations and local conservation of energy-momentum tensor can be found to be given by,
\begin{eqnarray}
    \frac{dH}{dt} &=& -H^2 - {\frac{2}{3}}\Sigma^2 - {\frac{1}{6}}(\rho + 2p_a + p_b)\,, \nonumber \\
    \frac{d\Sigma}{dt} &=& -3H\Sigma + \frac{1}{\sqrt{3}}(p_a - p_b)\,, \\
    \frac{d\rho}{dt} &=& -3H(\rho + \frac{2p_a + p_b}{3}) - \frac{2\Sigma}{\sqrt{3}}(p_a - p_b)\,, \nonumber
\label{eq:b1efe-new}
\end{eqnarray}
along with the constraint equation,
\begin{equation}
    H^2 = \frac{\rho}{3} + \frac{\Sigma^2}{3}\,.
\label{eq:b1consteqn}
\end{equation}
in units of $c=8 \pi G=1$.

Using the above constraint equation, we can define the dimensionless energy density parameter $\Omega=\rho/(3H^2)$ and the expansion-normalized dimensionless shear parameter $\sigma=\Sigma/(\sqrt{3}H)$. It is important to note that the energy density $\rho$ in Eq.~(\ref{eq:b1consteqn}), and consequently $\Omega$, comprises two components : the isotropic (ordinary and dark) matter denoted by $\Omega_m$, and the anisotropic dark energy denoted by $\Omega_{\rm ade}$. By defining the critical energy density as $\rho_{c,0}$ or simply $\rho_0=3H^2_0$, where $H_0$ represents the current value of the mean Hubble parameter (i.e., $H(t_0)=H_0$), we can express the constraint equation at present as
\begin{equation}
\Omega_{m,0}+\Omega_{\rm ade,0} + \sigma^2_0 = 1\,,
\label{eq:b1-constraint-eqn}
\end{equation}
where $\Omega_{m,0} = \rho_{m,0}/\rho_0$ and $\Omega_{\rm ade,0} = \rho_{\rm ade,0}/\rho_0$ that represent the fractional energy densities due to respective sources as indicated at the current time `$t_0$'. Hereafter, we will omit the subscript `$0$' for brevity, denoting them simply as $\Omega_{m}$ and $\Omega_{\rm ade}$. However, their usage should be clear from the context, whether they refer to energy densities at the current time or time-dependent variables. Finally, we define a dimensionless time variable as $\tau=\ln(A)$, leading to $H=d\tau/dt$.

The relation between direction-dependent luminosity distance ($d_L$) and redshift ($z$) of a Type Ia Supernova object observed in the direction $\hat{n}=(\theta,\phi$) in a Bianchi-I universe is given by~\cite{koivisto2008},
\begin{equation}
 d_L(\hat{n}) = c(1 + z_{\rm hel}) \int_{A(z^*)}^1 \frac{dA}{{A^2}H} \frac{(1 - e^2)^{1/6}}{(1 - e^2\cos^2\zeta)^{1/2}}\,.
\label{eq:b1lumdist}
\end{equation}
Here, $\zeta = \cos^{-1}(\hat{\lambda}\cdot\hat{n})$ represents the angle between the cosmic preferred axis $\hat{\lambda}$' (i.e., the $z$-axis of Eq.~\ref{eq:b1metric-plnr}) and the location of a Type Ia Supernova at `$\hat{n}$' in the sky. Further, `$z_{\rm hel}$' and `$z^*$' denotes redshift of a Type Ia Supernova in heliocentric and cosmic (CMB) frames respectively~\cite{davis2011}.

To solve for the integral in the above expression for luminosity distance `$d_L$', one must alongside solve the complete set of evolution equations that are setup in terms of various dimensionless variables defined earlier, following Einstein's field equations and others. They can be found to be~\cite{aluri2013sn1a,anshul2024},
\begin{eqnarray}
\label{eq:b1evoleq}
    \frac{h'}{h} &=& -\frac{3}{2}(1+\sigma^2 +\bar{w}\Omega_{\rm ade})\,, \nonumber\\
    (e^2)' &=& 6\sigma(1-e^2)\,, \nonumber \\
    \sigma' &=& -\frac{3}{2}[\sigma(1-\sigma^2) - (\frac{2}{3}\delta_w + \bar{w}\sigma)\Omega_{\rm ade}]\,, \\
    \Omega_{\rm ade}' &=& -3\Omega_{\rm ade}(\bar{w} + \frac{2}{3}{\delta_w}\sigma - \bar{w}\Omega_{\rm ade} - \sigma^2)\,, \nonumber \\
    \Omega_{m}' &=& 3\Omega_{m}(\bar{w}\Omega_{\rm ade} + \sigma^2)\,. \nonumber
\end{eqnarray}
Here, $\bar{w}$ denotes average equation of state parameter given by $\bar{w}=(2w_a+w_b)/3$, and their difference as $\delta_w=w_a-w_b$. The parameter `$h$' in the preceding equations stands for the dimensionless Hubble parameter in units of $100~\text{km/s/Mpc}$ i.e., $h=H/(100~\text{km/s/Mpc})$. Finally, for any of the parameters in the evolution equations given above, say $\theta$, a derivative with respect to dimensionless time variable `$\tau$' is denoted by $\theta'$.

Thus the cosmological parameters to be constrained in our model are 
$
\{h_0, e^2_0, \sigma_0, \Omega_{\rm ade}, \Omega_{m}, \hat{\lambda} = (l_a,b_a)\}
$
along with $w_a$ and $w_b$, or equivalently $\bar{w}$ and $\delta_{w}$. We consider four cases of anisotropic equation of state parameterization for the dark energy as described below :
\begin{description}
  \item[Case-I] `$w_a$' as free parameter, and $w_b = -1$ (fixed),
  \item[Case-II] $w_a = -1$ (fixed), and `$w_b$' as free parameter,
  \item[Case-III] Both $\bar{w}$ and $\delta_{w}$ as free parameters.
  \item[Case-IV]  $w_a = w_b = -1$ as the simplest Bianchi-I extension of $\Lambda$CDM.
\end{description}
Following the constraint equation of Eq.~(\ref{eq:b1-constraint-eqn}), $\Omega_{m}$ is treated as derived parameter, while $\Omega_{\rm ade}$ and $\sigma_0$ are varied freely to obtain constraints. Further, the cosmic anisotropy axis in our model, denoted by $\hat{\lambda} = (l_a,b_a)$, will be estimated in galactic coordinates.

We compare our results with the standard flat $\Lambda$CDM model, for which the luminosity distance in given by
\begin{equation}
d_L = \frac{c(1 + z_{\rm hel})}{H_0}\int_0^{z^*}{\frac{dz}{\sqrt{\Omega_{m}(1 + z)^3 + \Omega_\Lambda}}}\,,
\label{eq:lcdm-lumdist}
\end{equation}
where $\Omega_{m}$ represents the current fraction of matter density (comprising dust-like dark+visible matter) in the universe, and $\Omega_\Lambda$ is the dark energy density fraction. The constraint equation relating these two parameters at current times is $\Omega_m+\Omega_\Lambda=1$. Hence, within the standard concordance model, we have two cosmological parameters to be constrained: $H_0$ (or equivalently $h_0$) and $\Omega_{\Lambda}$ (treating $\Omega_m$ as a derived parameter i.e., $\Omega_m=1-\Omega_\Lambda$), with SNIa absolute magnitude $M_0$ remaining as a nuisance parameter. This standard model serves as our reference for comparison with rest of the models that we study here.

As an extension to the flat $\Lambda$CDM model, we also obtain constraints on the $w$CDM model in which the luminosity distance is given by
\begin{equation}
d_L = \frac{c(1 + z_{\rm hel})}{H_0}\int_0^{z^*}{\frac{dz}{\sqrt{\Omega_{m}(1 + z)^3 + \Omega_{\rm de}(1+z)^\eta}}}\,,
\label{eq:wcdm-lumdist}
\end{equation}
where $\eta=3(1+w_{\rm de})$ and $w_{\rm de}$ is the dark energy equation of state that could be different from ``$-1$''.
Our parameter set in this case is $\{ h_0, \Omega_{\rm de}, w_{\rm de}, M_0 \}$ i.e., we choose to constrain $\Omega_{\rm de}$, and the matter density fraction is given by $\Omega_m=1-\Omega_{\rm de}$ following the first Friedmann equation in a flat FLRW space-time.

\begin{figure}
    \centering
    \includegraphics[width=0.44\textwidth]{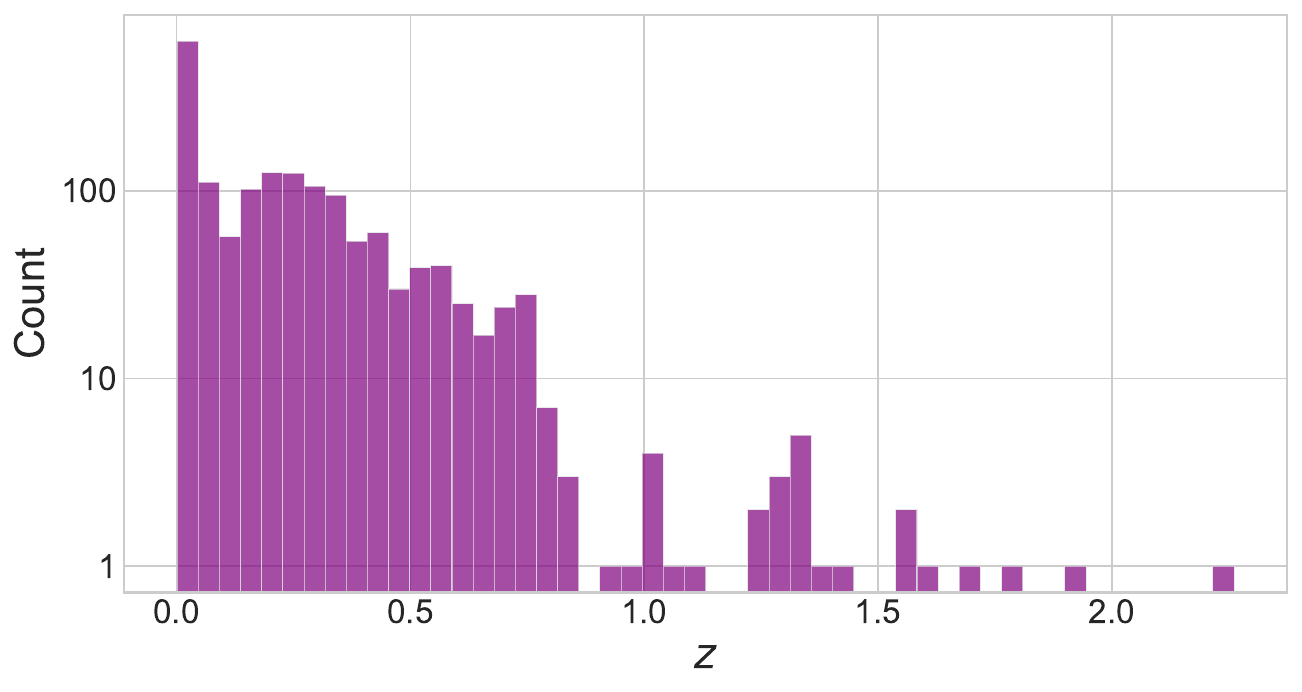}
    ~
    \includegraphics[width=0.44\textwidth]{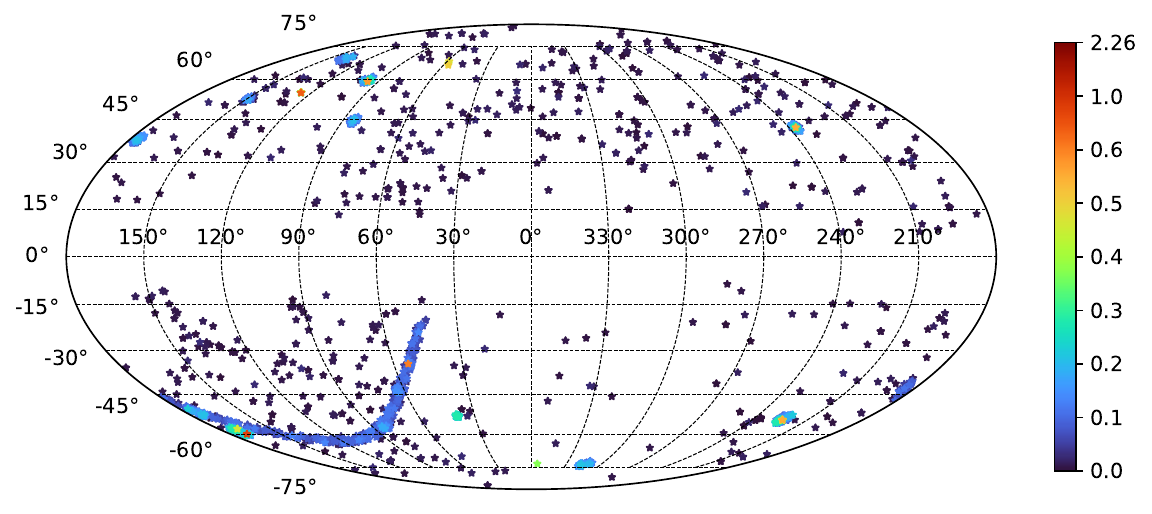}
    \caption{\emph{Top :} Redshift distribution of SNIa in Pantheon+ data in 100 bins.
    		 \emph{Bottom :} Sky distribution of Pantheon+ SNIa in galactic coordinates.}
    \label{fig:pantheon+_dist}
\end{figure}

\section{Data set and Likelihood analysis}
\label{sec:SNIa-data-method}

The Pantheon+ compilation of Type Ia Supernovae (SNIa)\footnote{\url{https://github.com/PantheonPlusSH0ES/DataRelease}} comprises of 1701 light curves in turn containing approximately 1550 unique spectroscopically confirmed SNIa~\cite{scolnic2022}. We note that the light curves of the remaining SNIa, i.e., those SNIa other than the approximately 1550 out of 1701, are either same Supernova (SN) observed by different surveys or what are called ``SN siblings'' that mean Supernovae (SNe) found in the same host galaxy. Notable enhancements in the latest data set include the addition of new SNIa especially at low redshifts, as well as adjustments for peculiar velocity and host galaxy bias in the SNIa covariance matrix.
The redshift and spatial (sky) distribution of SNIa in Pantheon+ compilation are shown in the \emph{top} and \emph{bottom} panels of Fig.~[\ref{fig:pantheon+_dist}] respectively.

Now, we outline the likelihood function ($\mathcal{L}$), that depends on various fitting parameters, including the cosmological parameters under investigation. Using an {\rm MCMC} method, we maximize the likelihood function by sampling different parameters that it depends on.

Central to deriving constraints on various (cosmological) parameters of interest using SNIa is the theoretical relation between redshift ($z^*$) and the luminosity distance ($d_L$) that depends on different cosmological parameters via the distance modulus ($\mu^{\rm th}$) that is given by,
\begin{equation} 
\mu^{\rm th} = 5\log_{10}\left( \frac{d_L({\boldsymbol\theta},z^*)}{10~\text{pc}} \right) = m_B-M_0\,.
\label{eq:distmod}
\end{equation}
Here, $m_B$ represents the apparent $B$-band magnitude of an SNIa, while $M_0$ denotes its absolute magnitude. Conventionally, `$M_0$' is assumed to be same for all SNIa, and we follow the same here also treating it as a global free parameter to be fitted. Notably, $M_0$ exhibits a strong degeneracy with the Hubble parameter $H_0$ and is pivotal in motivating the utilization of SH0ES Cepheid variables~\cite{riess2022}.
 
Our $\chi^2({\boldsymbol \theta})$ function ($=-2\ln(\mathcal{L}(\boldsymbol \theta))$) that depends on all the parameters being fit `${\boldsymbol \theta}$' is defined as,
\begin{equation}
\chi^2({\boldsymbol \theta}) = \Delta{\boldsymbol \mu}^T ({{\bf C}_{\rm stat+sys}^{\rm SN} + {\bf C}_{\rm stat+sys}^{\rm Cepheid}})^{-1} \Delta{\boldsymbol \mu}\,,
\label{eq:chi2-cov}
\end{equation}
which is minimized to obtain relevant constraints.
Here, $\Delta{\boldsymbol \mu}=(\Delta \mu_1, \Delta \mu_2, \hdots, \Delta \mu_n)^T$ is a column vector of size $1\times n$ and `$n$' is the number of SNIa in the data set we are using. Then the residual $\Delta \mu_i=\mu^{\rm obs}_i - \mu^{\rm th}_i$ for an $i$th SNIa is defined as,
\begin{eqnarray}
    \Delta \mu_i = \hspace{20em} \\
    \left\{
\begin{array}{l l}
    m^i_{B,\rm corr} - M_0 - \mu_i^{\rm Cepheid}, & \text{if } i \in \text{Ceph. hosts}\,, \\
    m^i_{B,\rm corr} - M_0 - \mu^{\rm th}({\boldsymbol\theta},z^*_i), & \text{otherwise}\,.
\end{array} \nonumber
\right.
\end{eqnarray}
where,
\begin{equation}
m^i_{B,\rm corr}=m^i_{B} + \alpha x_{1,i} - \beta \texttt{c}_i + \gamma G_{{\rm host},i}\,,
\label{eq:mbcorr}    
\end{equation}
is the corrected apparent magnitude of an SNIa. The observed peak $B$-band magnitude `$m_B$', light curve width parameter `$x_1$', and color parameter `\texttt{c}' are determined during SNIa light curve fitting. Then, `$G_{\rm host}$' accounts for host galaxy stellar mass ($M_{\rm host}$) dependence. It is defined as a step function based on the value of $M_{\rm host}$ as, 
\begin{equation}
G_{{\rm host},i} = 
\begin{cases} 
+\frac{1}{2} & \text{if } M_{{\rm host},i} > 10^{10} M_{\odot}, \\
-\frac{1}{2} & \text{if } M_{{\rm host},i} < 10^{10} M_{\odot},
\end{cases}
\end{equation}
where \(M_{\odot}\) represents the solar mass. The values of \(m_B\), \(x_1\), \(\texttt{c}\), \(G_{{\rm host},i}\), and \(\mu_i^{\rm Ceph}\) are provided in the Pantheon+ dataset.

Therefore, $\mu^{\rm obs}_i=m^i_{B,\rm corr} - M_0$, and $\mu^{\rm th}_i=\mu^{\rm th}({\boldsymbol\theta},z^*_i)$ is the theoretical distance modulus that depends on model parameters being constrained, while $z^*_i$ is the redshift of that Supernova. For any SNIa whose host galaxy also contains a Cepheid variable, $\mu^{\rm th}_i$ is taken to be $\mu_i^{\rm Cepheid}$ itself that is well measured. In addition to $M_0$, the coefficients $\alpha$, $\beta$, and $\gamma$ are also fit as global free parameters in deriving the constraints.

Considering the correlations between SNIa that are calibrated simultaneously due to both statistical and systematic uncertainties in SNIa light curve fitting, we also use the full covariance matrix.
${\bf C}_{\rm stat+sys}^{\rm SN}$ denotes the SN covariance matrix and ${\bf C}_{\rm stat+sys}^{\rm Cepheid}$ contains the SH0ES Cepheid host-distance covariance matrix, both of which are provided as part of Pantheon+ data set.

Constraints will be derived for the three cases {I}, {II} and {III} of anisotropic equation of state parameterization as described in the previous section. We also obtain constraints on case {IV} where a Bianchi-I background with isotropic sources is considered, and then the $w$CDM and standard flat $\Lambda$CDM models for completeness.

\section{Results}
\label{sec:results}

\subsection{Cosmological constraints}
\label{sec:cosmo-par-constr}
We discuss the results obtained from cosmological parameter fitting of Bianchi-I model for our universe with anisotropic dark energy in this section. For the Case-I, $w_b = -1$ is fixed i.e., equation of state for dark energy along the anisotropy axis is fixed and the total free parameters for MCMC fitting are $\{h_0, e^2_0, \sigma_0, \Omega_{\rm ade}, w_a, (l_a,b_a)\}$. For the Case-II, $w_a = -1$ is fixed i.e., equation of state for dark energy in the plane normal to the axis of anisotropy is fixed and the free parameters are now $\{h_0, e^2_0, \sigma_0, \Omega_{\rm ade}, w_b, (l_a,b_a)\}$. For the Case-III, both the parameters $\bar{w}=(2w_a+w_b)/3$ and $\delta_w=w_a-w_b$ are kept as free parameters. Hence the total free parameters in Case-III are $\{h_0, e^2_0, \sigma_0, \Omega_{\rm ade}, \bar{w}, \delta_w, (l_a,b_a)\}$.

As Case-IV, we also consider a Bianchi-I universe with isotropic sources i.e., $w_a=w_b=-1$ (fixed) for cosmological constant ($\Lambda$) as dark energy, and $w_a=w_b=0$ for matter. Hence the total cosmological parameters to constrain are $\{h_0, e^2_0, \sigma_0, \Omega_\Lambda, (l_a,b_a)\}$.  Note that the fractional matter density $\Omega_m$ in all these four cases is a derived parameter following the constraint equation given by Eq.~(\ref{eq:b1-constraint-eqn}). (However, in Case-IV, the constraint equation is $\Omega_m+\Omega_\Lambda+\sigma_0^2=1$.)

\begin{figure*}
\centering
\includegraphics[width=0.45\textwidth]{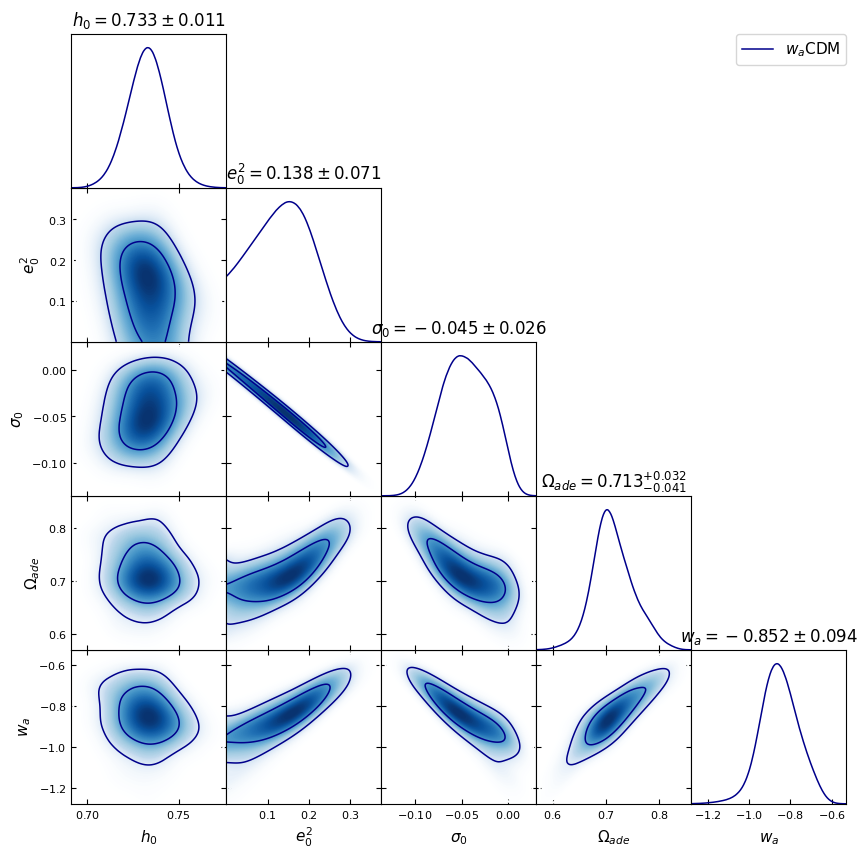}
% ~Figures/ade
\includegraphics[width=0.45\textwidth]{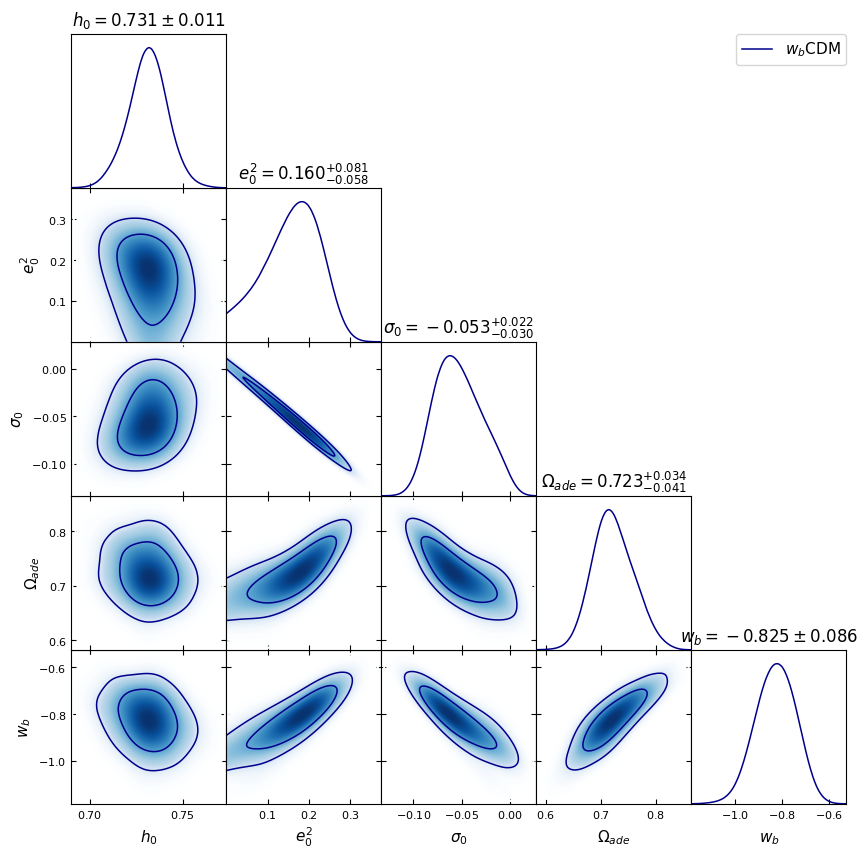}
\includegraphics[width=0.45\textwidth]{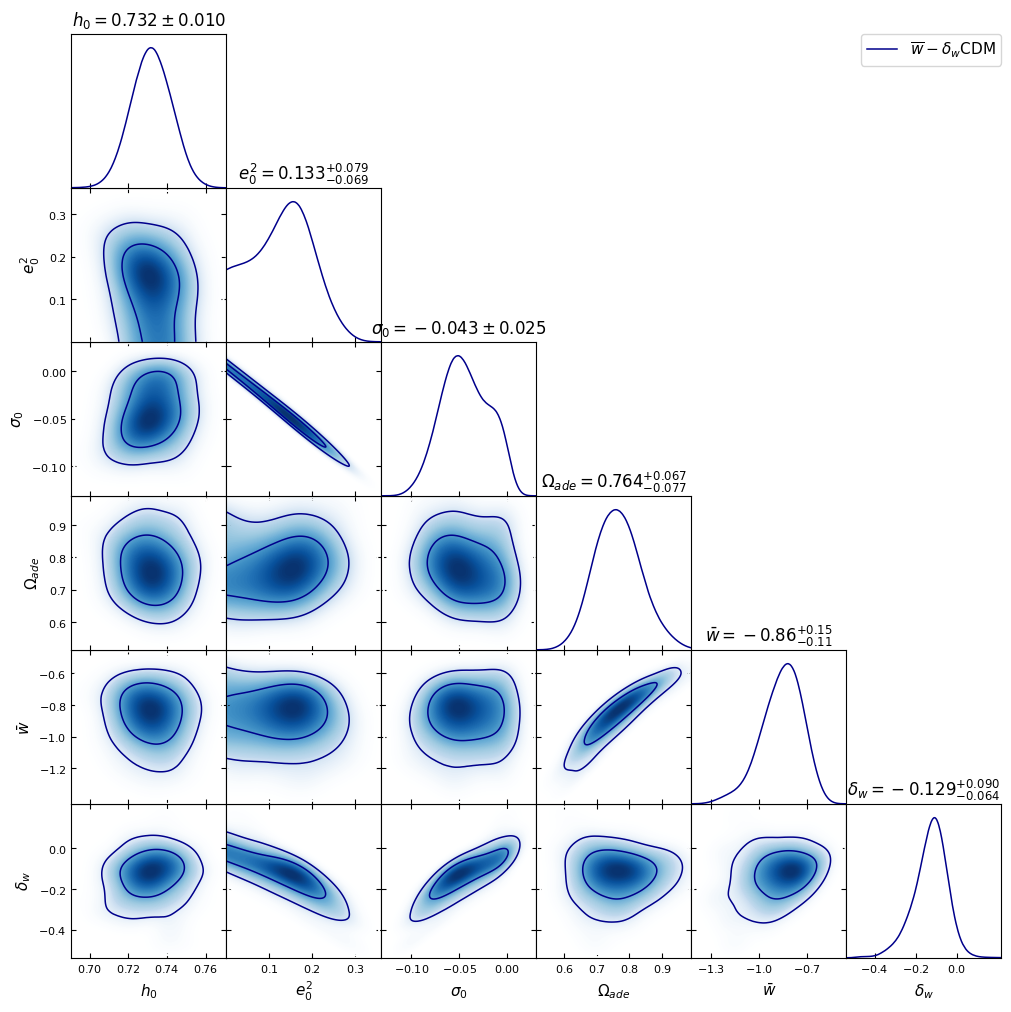}
\includegraphics[width=0.45\textwidth]{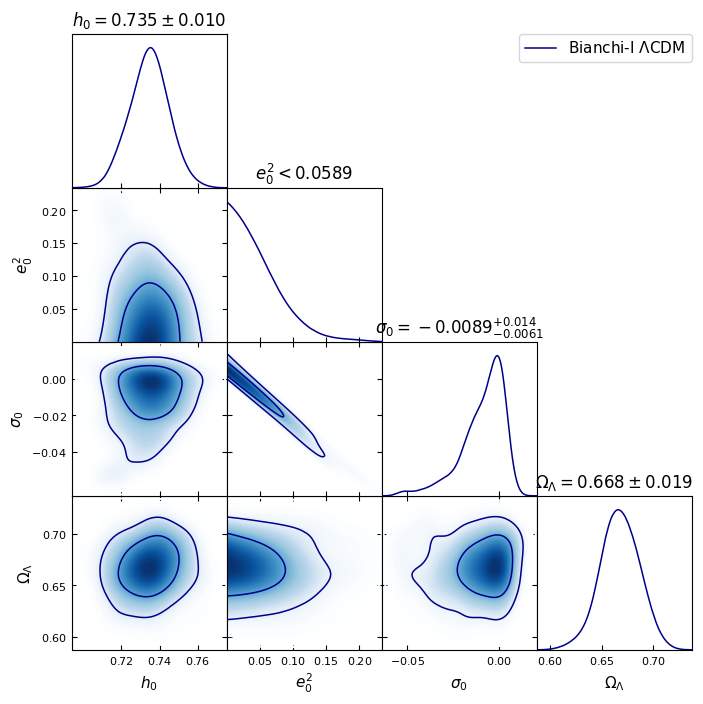}
\caption{1$\sigma$ and 2$\sigma$ CL constraints on all cosmological parameters in the four cases of an axisymmetric Bianchi-I universe with different equations of state parameterization of dark energy. \emph{Top left :} $w_a$CDM model where e.o.s. of dark energy in the residual plane of symmetry viz., $w_a$ is freely varied, fixing $w_b=-1$. \emph{Top right :} $w_b$CDM model where dark energy e.o.s. along cosmic $z$-axis viz., $w_b$ was free to vary, while fixing $w_a=-1$.
\emph{Bottom left :} $\bar{w}-\delta_w$CDM model where the free parameters are the mean e.o.s. of anisotropic dark energy, $\bar{w}=(2w_a+w_b)/3$, and the difference in the two equations of state $\delta_w=w_a-w_b$ that captures anisotropic expansion in different directions.
\emph{Bottom right :} B-I~$\Lambda$CDM model where an anisotropic extension of standard $\Lambda$CDM model i.e., $\Lambda$CDM sources in a Bianchi-I background was studied. 
}
\label{fig:model-b1-case1234}
\end{figure*}

In Fig.~[\ref{fig:model-b1-case1234}], we present two-dimensional contour plots illustrating the $1\sigma$ and $2\sigma$ confidence levels (CL) for various cosmological parameters of Case-I, Case-II, Case-III and Case-IV viz., ${h_0, e^2_0, \sigma_0, \Omega_{\rm ade}}/\Omega_\Lambda$ and respective anisotropic dark energy equation of state parameters depending on the model.

\begin{figure*}
\centering
\includegraphics[width=0.43\textwidth]{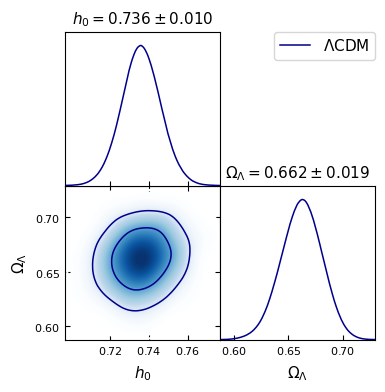}
% ~Figures/frw
\includegraphics[width=0.44\textwidth]{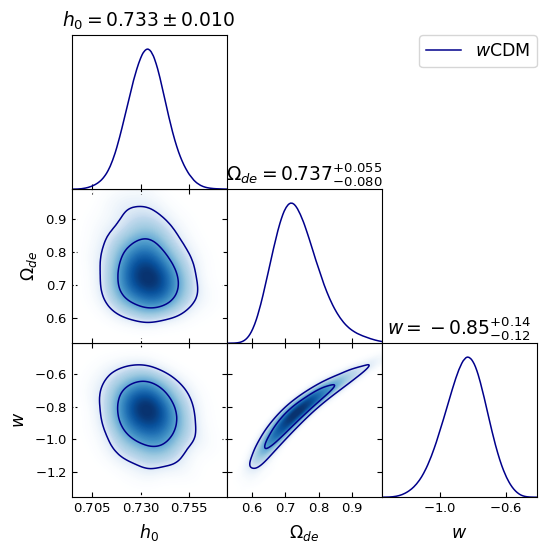}
\caption{Two dimensional contour plots of various cosmological parameters with 1$\sigma$ and 2$\sigma$ CL in our base $\Lambda${CDM}(\emph{left}), and the $w${CDM} (\emph{right}) cosmological model.}
\label{fig:model-lcdm-wcdm}
\end{figure*}

For the $\Lambda$CDM and $w$CDM models, free parameters are $\{h_0,\Omega_{\Lambda}\}$ and $\{h_0,\Omega_{\rm de},w\}$ respectively. $1\sigma$ and $2\sigma$ confidence levels (CL) for these parameters are shown as a corner plot in Fig.~[\ref{fig:model-lcdm-wcdm}] in the \emph{left} and \emph{right} panels respectively.
Here also, following the respective constraint equations involving energy density fractions, we estimate $\Omega_\Lambda$ or $\Omega_{\rm de}$ as a free parameter thus calculating $\Omega_m$ as a derived parameter.

\begin{figure*}
\centering
\includegraphics[width=0.85	\textwidth]{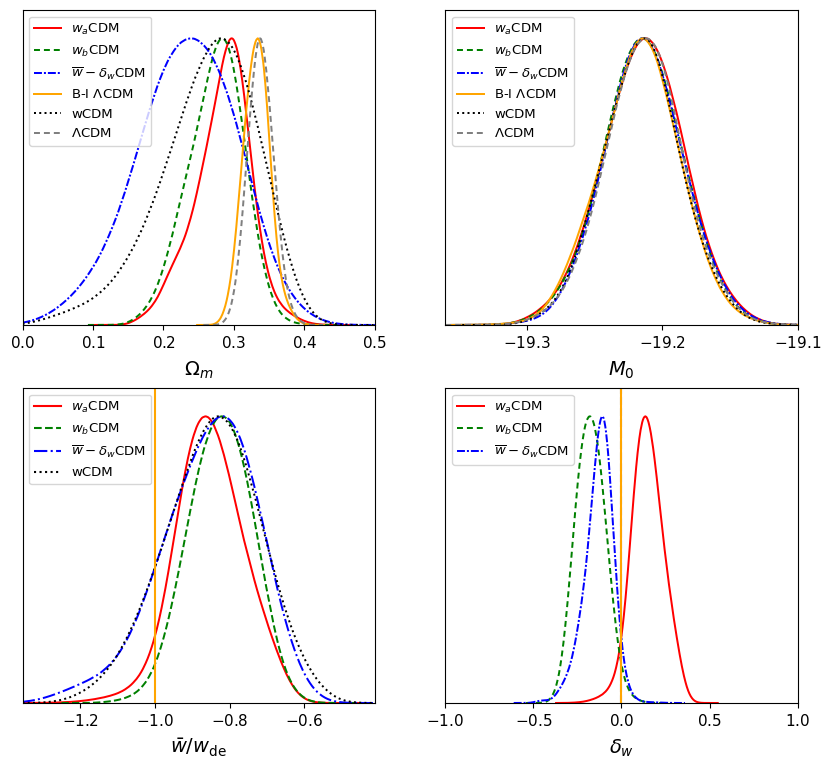}
\caption{Posterior distribution of (the derived parameter) isotropic matter density `$\Omega_m$' (\emph{top left}), absolute $B$-band magnitude of Type Ia Supernova `$M_0$' (\emph{top right}), various dark energy equation of state parameters $\bar{w}$/$w_{\rm de}$ (\emph{bottom left}), and the difference in dark energy equation of state $\delta_w = w_a - w_b$ (\emph{bottom right}) for all cases of Bianchi-I background as well as $w$CDM model.}
\label{fig:Om_M0_w_dw_all}
\end{figure*}

For all models that we studied, posterior distributions of $\Omega_m$ (\emph{top left}) and $M_0$ (\emph{top right}), and also $\bar{w}$ and $\delta_w$ (\emph{bottom panels} for easy comparison with base model) are presented in Fig.~[\ref{fig:Om_M0_w_dw_all}].
All the parameter constraints for the six models viz., four cases of axisymmetric Bianchi-I universe (Case-I, II, III and IV), and the flat $\Lambda$CDM and $w$CDM models are presented in  Table~{\ref{tab:par-val-mcmc}} and \ref{tab:par-val-mcmc-lcdm}.

Interestingly, the dark energy density fraction is found to be lowest for B-I $\Lambda$CDM model compared to rest of the models studied here including the standard flat $\Lambda$CDM and $w$CDM models. Further the cosmic shear parameter, $\sigma_0$, and the eccentricity parameter, $e^2_0$, were also lower for that model.
We could only get an upper bound for $e^2_0$ parameter in the B-I $\Lambda$CDM model. This indicates that $a_0=b_0=1$ roughly, meaning that their relative scaling is not different from `1' at current times in that model. This may perhaps be understood in the following way : even though the background is anisotropic, in the presence of isotropic sources the universe isotropizes faster than when anisotropic sources are present.

We note that the estimated cosmic shear parameter, $\sigma_0$, is preferring a negative value in all the four anisotropic Bianchi-I cases, irrespective of its actual value.
Given the fit for non-zero cosmic shear from SNIa data in all four cases, we may infer that the cosmic expansion is perhaps different in different directions.

Here we briefly comment about the implications of cosmic shear on the evolutionary history of our universe. Deviation from isotropic expansion in a Bianchi-I model compared to the FLRW background is captured by the shear parameter `$\sigma$'. In a Bianchi-I model with isotropic sources, we can analytically derive the variation of shear energy density with respect to `$A$'. It is given by $\Omega_{\sigma}=\Omega_{\sigma,0}/A^6$, where $\Omega_\sigma=\sigma^2$ (and thus $\Omega_{\sigma,0}=\sigma_0^2$). Hence it grows faster than the radiation energy density fraction viz., $\Omega_r = \Omega_{r,0} /A^4$ as we go back in time.
Therefore it could even dominate the radiation density fraction at early times depending on its current value (as estimated), contradicting observations.
Level of anisotropy in the late universe assuming a Bianchi background are discussed earlier in the literature. Constraints using SNIa and $H(z)$ data (upto $z \lesssim 2.4$) on the shear density parameter $\Omega_{\sigma,0}$, were found to be $\mathcal{O}(10^{-3})$ \cite{2011campanelli,Wang:2018,Akarsu:2019,Amirhashchi:2020,2023Akarsu}.
However, such a value (as also found to be so in the present work) is high and would spoil the standard cosmology for $z\geq10$.

\begin{figure*}
\centering
\includegraphics[height=5.9cm]{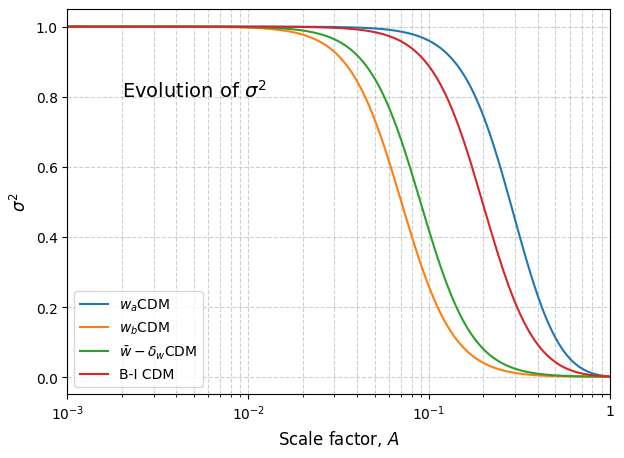}
~
\includegraphics[height=5.9cm]{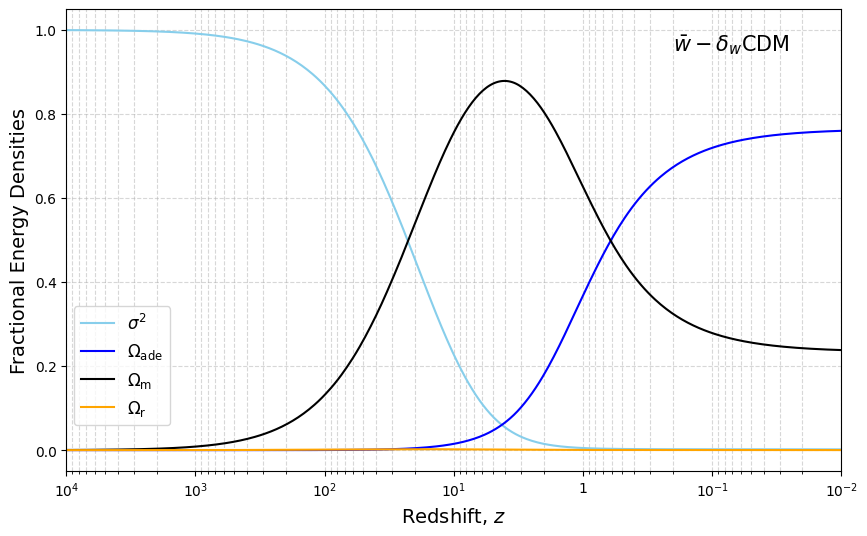}
\caption{{\emph{Left :} Evolution of shear energy density $\Omega_\sigma=\sigma^2$ with mean scale factor `$A$'.
\emph{Right :} Evolution of energy density fraction of various sources, as indicated, in
a $\bar{w}-\delta_w$CDM model with redshift.
This evolution is as per the cosmological constraints derived from the data (presented in Table~\ref{tab:par-val-mcmc} and \ref{tab:par-val-mcmc-lcdm}). See text for more discussion on shear density parameter.}}
\label{fig:shear-wdwcdm}
\end{figure*}

In \emph{left} panel of Fig.~[\ref{fig:shear-wdwcdm}], we show the evolution of shear energy density fraction, $\Omega_\sigma=\sigma^2$ for various cases of anisotropic dark energy equation of state parameterizations that we studied here.  As mentioned earlier, shear varies as $\Omega_{\sigma}\propto1/A^6$ in case of B-I~$\Lambda$CDM where isotropic matter and cosmological constant as dark energy were considered as sources in a Bianchi-I background. One can clearly see a difference in its evolution from this behaviour - slower or faster decay - towards isotropization at current times. Same initial value for the `residual' shear at present was used to illustrate this difference viz., $\sigma_0=-0.043$ as estimated in $\bar{w}-\delta_w$CDM model.

One also notices from the same figure that shear energy density dominates at early times in all cases if its true value is as we estimated. This contradicts the constraints  we have from early universe observations. To illustrate it further, we evolve the coupled equations in Eq.~(\ref{eq:b1evoleq}) back in time (specifically redshift) using the constraints that we derived for various energy density fractions in Case-III i.e., $\bar{w}-\delta_w$CDM model as initial values. The results are presented in the \emph{right} panel of Fig.~[\ref{fig:shear-wdwcdm}]. It is clear from the figure that with $\Omega_{\sigma,0}\sim10^{-3}$, shear dominates at early times (high redshifts), even dominating the energy density due to radiation at $z\sim1100$ contradicting CMB observations.

Using early universe probes viz., Cosmic Microwave Background (CMB) (formed at $z \sim 1100$ in the standard model) and Big Bang Nucleosynthesis (BBN) ($z\sim10^9$) data gives much tighter constraints on the shear parameter as $\Omega_{\sigma,0} \sim \mathcal{O}(10^{-23})$, for example, as argued in Refs.~\cite{Barrow:1976,Campanelli2011He4,Akarsu:2019}. Recently, in Ref.~\cite{saadeh2016MNRAS}, an upper bound on cosmic shear was found to be $|\sigma_0|\lesssim \mathcal{O}(10^{-10})$ due to scalar modes (upper bounds on shear due to vector and tensor modes were also reported by them that are $\lesssim \mathcal{O}(10^{-7})$).
An isotropization condition for the shear parameter and its asymptotic limits for very small values of mean scale factor ($A \ll 1$) were discussed earlier in Refs.~\cite{Campanelli2011ade,Tedesco2018}.

\begin{figure*}
    \centering
    \includegraphics[scale=0.75]{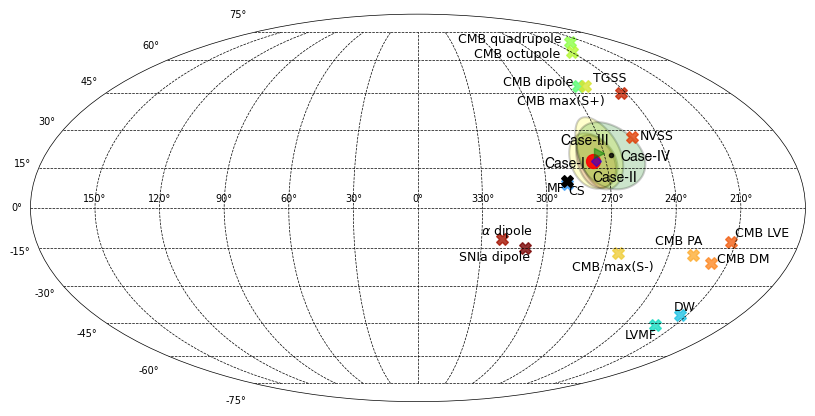}
    \caption{
    Mollweide projection of the cosmic preferred axis, with $1\sigma$
    confidence contours, found using Pantheon+ SNIa data for the three cases
    of anisotropic equation of state parameterization of dark energy (Case-I,
    II and III) and the B-I~$\Lambda$CDM model (Case-IV) which is an anisotropic
    Bianchi-I extension of the base $\Lambda$CDM model.
    Also shown are similar axis of anisotropy found in various
    astronomical/cosmological data :
    CMB dipole~\cite{plk2018maps}, CMB quadrupole and octopole axes~\cite{plk2013isostat},
    CMB Even and odd mirror parity axes ($S_\pm$), CMB Hemispherical Power Asymmetry (HPA)
    axis, Dipole Modulation (DM) axis~\cite{plk2015isostat}, CMB HPA axis from Local Variance
    estimator (LVE)~\cite{akrami2014}, Radio dipole in NVSS and TGSS data~\cite{bengaly2018},
    Fine structure constant ($\alpha$) dipole and SNIa Dipole~\cite{alpha_dipole}, and Cosmic
    strings (CS), Domain walls (DW), Magnetic fields (MF) and Lorentz Violation generated
    magnetic fields (LVMF) as sources of anisotropy~\cite{anshul2024}.}
    \label{fig:cosmic_axes}
\end{figure*}

One of the most interesting results from this work is constraints derived on cosmic preferred axis due to anisotropic expansion of the universe all of which align in the same direction.
The constraints on the anisotropy axis in all four cases of Bianchi-I model are: $(l_a, b_a) \approx (276^\circ, 18^\circ)$, $(275^\circ, 18^\circ)$, $(272^\circ, 21^\circ)$ and $(267^\circ, 20^\circ)$ respectively in galactic coordinates. These axes are shown in Fig.~[\ref{fig:cosmic_axes}] along with their $1\sigma$ confidence contours.
All these axes are very close by with nearly same errorbars.
More interestingly it aligns with the preferred axis found in our previous work~\cite{anshul2024} where anisotropic matter sources were assumed instead of anisotropic dark energy, specifically when cosmic strings (CS) and magnetic fields (MF) were taken to be the source of anisotropy. There we found the anisotropy axis to be $(l_a, b_a) \approx (290^\circ, 9^\circ)$ and $(290^\circ, 10^\circ)$ with CS and MF respectively.
The anisotropy axes found here are depicted alongside notable anisotropy axes seen in various data sets, such as CMB, radio data, SNIa (from previous releases), the CMB dipole ($\ell = 1$), etc., as mentioned in the caption of Fig.~[\ref{fig:cosmic_axes}]. Here we note that in the previous work~\cite{anshul2024}, current energy density fractions for these two sources, namely cosmic strings and magnetic fields, were consistent with zero. However we cannot say whether this indicates a systematic in the data that is being picked up by the anisotropic parameterization of dark energy as source in an anisotropic background space-time.

For the anisotropic dark energy equation of state parameter, we get $w_a = -0.852\pm0.094$ and $w_b = -0.850\pm0.099$ for the first two cases of $w_a$CDM and $w_b$CDM, while we get
$\bar{w}=-0.86^{+0.15}_{-0.11}$ for the third case of $\bar{w}-\delta_w$CDM. All of these results are shown separately in the \emph{bottom panels} of Fig.~[\ref{fig:Om_M0_w_dw_all}] in terms of $\bar{w}$ and $\delta_w$ for easy comparison with the $\Lambda$CDM and $w$CDM models. The equation of state for dark energy (cosmological constant) in the base $\Lambda$CDM model where $w_{\rm de} = -1$ and $\delta_w=0$ are depicted in those two bottom panels as vertical \emph{yellow} lines.
The difference between equation of state parameters in and normal to the plane of axisymmetry i.e $\delta_w = w_a - w_b$ were found to be $0.148\pm 0.094$, $-0.175\pm 0.086$, and $-0.129^{+0.090}_{-0.064}$ for the $w_a$CDM, $w_b$CDM and $\bar{w}-\delta_w$CDM models, respectively. From the posterior distributions shown in the bottom two panels of Fig.~[\ref{fig:Om_M0_w_dw_all}], we note an interesting observation that in $\bar{w}-\delta_w$CDM model (Case-III) where both $w_a$ and $w_b$ are freely varied, their estimates are closer to the values found for $w_b$CDM model (Case-II) where we fixed $w_a=-1$ but only varied $w_b$ freely. (We remind that in Case-III we actually sampled $\bar{w}$ and $\delta_w$ as free parameters.)

Finally, we also display constraints on the SNIa absolute magnitude `$M_0$' in the \emph{top right} panel of Fig.~[\ref{fig:Om_M0_w_dw_all}]. As is obvious from the plot, it is essentially same for all models that we studied which is $M_0 \approx -19.247\pm{0.029}$ and is in agreement with the value originally reported by Pantheon+ collaboration~\cite{scolnic2022}. These values are also listed in Table~\ref{tab:par-val-mcmc} and \ref{tab:par-val-mcmc-lcdm}, along with estimates of the coefficients of SNIa light curve stretch, color and host-galaxy stellar mass bias parameter viz., $\alpha$, $\beta$ and $\gamma$ in Eq.~\ref{eq:mbcorr}.

\begin{table*}
 \centering
    \begin{tabular}{| c || c | c | c | c |}
    \hline
    \rowcolor{LightCyan}
    Parameter & Case I ($w_a$CDM)  & Case II ($w_b$CDM)  & Case III ($\bar{w}-\delta_w$CDM)  & Case IV (B-I $\Lambda$CDM) \\
    \hline
    $h_0$  & $0.733\pm 0.011$ & $0.731\pm 0.011$ & $0.732\pm 0.010$ & $0.735\pm 0.010$ \\[2pt]

    \rowcolor{LightCyan}
    $\sigma_0$ & $-0.045\pm0.026$ & $-0.053^{+0.022}_{-0.030}$ & $-0.043\pm 0.025$ & $-0.0089^{+0.014}_{-0.0061}$ \\[2pt]
    
    $e^2_0$ & $0.138\pm 0.071$ & $0.160^{+0.081}_{-0.058}$ & $0.133^{+0.079}_{-0.069}$ & $<0.0589$ \\[2pt]

    \rowcolor{LightCyan}
    $\Omega_{\rm de}$ & $0.713^{+0.032}_{-0.041}$ & $0.723^{+0.034}_{-0.041}$ & $0.764^{+0.067}_{-0.077}$ & $0.668\pm 0.019$ \\[2pt]
        
    $^*\Omega_m$ & $0.284^{+0.044}_{-0.032}$ & $0.274^{+0.044}_{-0.035}$ & $0.233^{+0.078}_{-0.067}$ & $0.332\pm 0.019$ \\[2pt]

    \rowcolor{LightCyan}
    $\bar{w}$ / $w_{\rm de}$  & $-0.951\pm 0.031^*$ & $-0.942\pm 0.029^*$ & $-0.86^{+0.15}_{-0.11}$ & $-1$ (fixed) \\[2pt]
    
    $\delta_{w}$  & $0.148\pm 0.094^*$ & $-0.175\pm 0.086^*$ & $-0.129^{+0.090}_{-0.064}$ & 0 (fixed) \\[2pt]

    \rowcolor{LightCyan}
    $w_a$  & $-0.852\pm 0.094$ & $-1$ (fixed) & $-0.90^{+0.16*}_{-0.12}$ & $-1$ (fixed) \\[2pt]
    
    $w_b$  & $-1$ (fixed) & $-0.850\pm0.099$ & $-0.81^{+0.17*}_{-0.11}$ & $-1$ (fixed)  \\[2pt]

    \rowcolor{LightCyan}
    $\hat{\lambda}=(l_a,b_a)$ & $(276.0^{+9.3}_{-13},17.9^{+7.3}_{-14})$ & $(274.7^{+9.8}_{-8.3},17.7^{+6.4}_{-14})$ & $(272^{+11}_{-9.8},21^{+8}_{-20})$ & $(267^{+19}_{-13},20^{+11}_{-15})$ \\[2pt]
    
    $M_0$  & $-19.214\pm 0.031$ & $-19.215\pm 0.030$ & $-19.213\pm 0.030$ & $-19.216\pm0.030$ \\[2pt]

    \rowcolor{LightCyan}
	$\alpha$ & $0.138 \pm 0.004$ & $0.138 \pm 0.004$ & $0.138 \pm 0.004$ & $0.137 \pm 0.004$ \\ [2pt]
	$\beta$ & $2.439 \pm 0.075$ &  $2.441 \pm 0.074$ & $2.433 \pm 0.075$ & $2.421 \pm 0.100$ \\ [2pt]

    \rowcolor{LightCyan}
	$\gamma$ & $0.037 \pm 0.010$ & $0.037 \pm 0.010$ & $0.036 \pm 0.010$ & $0.036 \pm 0.011$ \\ [2pt]
    \hline

	$k$ & 11 & 11 & 12 & 10 \\[2pt]
    
    \rowcolor{LightCyan}
    $\chi^2$ & 1592.78 & 1591.99 & 1592.06 & 1597.44 \\[2pt]
    
    $\Delta\chi^2$ & 9.83 & 10.62 & 10.55 & 5.18 \\[2pt]

    \rowcolor{LightCyan}
    $\Delta$ & 0.16 & -0.62 & 1.44 & 2.81 \\
    
    ${\rm AIC}_{\rm rel}$ & 0.92 & 1.36 & 0.48  & 0.24 \\

    \hline
    
    \end{tabular}
    \caption{Best fit values of all cosmological parameters of an anisotropic Bianchi-I
             model with three different anisotropic dark energy e.o.s.
             parameterizations (Case-I, II and III) and the usual cosmological constant
             as dark energy (Case-IV).
             `$\hat{\lambda}$' denotes axes of anisotropy estimated in each case in
             galactic coordinates, and `$k$' denotes the number of model parameters.
             $AIC_{\rm rel}$ and $\Delta$ enables a model selection criterion penalizing
             additional parameters introduced compared to the \emph{base} flat $\Lambda$CDM
             model.
             These constraints are derived using the latest Pantheon+ Type Ia Supernova
             dataset. The artesisk ($^*$) denotes derived parameters.}
\label{tab:par-val-mcmc}
\end{table*}

\begin{table}
 \centering
    \begin{tabular}{| c || c | c |}
    \hline
    \rowcolor{LightCyan}
    Parameter & $\Lambda$CDM & $w$CDM \\
    \hline
    $h_0$ & $0.736\pm 0.010$ & $0.733\pm 0.010$ \\[2pt]

    \rowcolor{LightCyan}
    $\Omega_{\rm de}$ & $0.736\pm 0.019$ & $0.737^{+0.055}_{0.080}$ \\[2pt]
        
    $^*\Omega_m$ & $0.338\pm 0.019$ & $0.263^{+0.080}_{-0.055}$ \\[2pt]

    \rowcolor{LightCyan}
    $w_{\rm de}$  & -1 (fixed) & $-0.85^{+0.14}_{-0.12}$ \\[2pt]
    
    $M_0$ & $-19.214\pm 0.029$ & $-19.215\pm 0.029$ \\[2pt]
    
    \rowcolor{LightCyan}
    $\alpha$ & $0.138 \pm 0.004$ & $0.137 \pm 0.004$ \\ [2pt]
    
    $\beta$ & $2.439 \pm 0.075$ & $2.434 \pm 0.073$ \\ [2pt]

    \rowcolor{LightCyan}    
    $\gamma$ & $0.036 \pm 0.010$ & $0.036 \pm 0.010$ \\

    \hline

	$k$ & 6 & 7 \\[2pt]
    
    \rowcolor{LightCyan}
    $\chi^2$ & 1602.62 & 1602.06 \\[2pt]
    
    $\Delta\chi^2$ & 0 & 0.56 \\[2pt]
   
    \rowcolor{LightCyan}
    $\Delta$ & 0 & 1.43 \\[2pt]

    AIC$_{\rm rel}$ & 1 & 0.48 \\
    \hline
    \end{tabular}
    \caption{Best fit cosmological parameters of the standard flat $\Lambda$CDM model and
    the $w$CDM model using  Pantheon+ SNIa dataset.
    $\Lambda$CDM model is taken as \emph{reference} model to compare all other models.
    See text for more details on model selection based on AIC. The artesisk ($^*$) denotes derived parameters.}
\label{tab:par-val-mcmc-lcdm}
\end{table}

\iffalse

\begin{table}[h]
\centering
\begin{tabular}{| l | c | c | c |}
\hline
\rowcolor{LightCyan}
Model & $\alpha$ & $\beta$ & $\gamma$ \\
\hline
$w_a$CDM & $0.138 \pm 0.004$ & $2.439 \pm 0.075$ & $0.037 \pm 0.010$ \\ [2pt]
\rowcolor{LightCyan}
$w_b$CDM & $0.138 \pm 0.004$ & $2.441 \pm 0.074$ & $0.037 \pm 0.010$ \\ [2pt]
$\overline{w}-\delta_{w}$CDM & $0.138 \pm 0.004$ & $2.433 \pm 0.075$ & $0.036 \pm 0.010$ \\[2pt]
\rowcolor{LightCyan}
B-I $\Lambda$CDM & $0.137 \pm 0.004$ & $2.421 \pm 0.100$ & $0.036 \pm 0.011$ \\ [2pt]
$w$CDM & $0.137 \pm 0.004$ & $2.434 \pm 0.073$ & $0.036 \pm 0.010$ \\ [2pt]
\rowcolor{LightCyan}
$\Lambda$CDM & $0.138 \pm 0.004$ & $2.439 \pm 0.075$ & $0.036 \pm 0.010$ \\
\hline
\end{tabular}
\caption{Marginalized 1D constraints on $\alpha$, $\beta$, and $\gamma$ with 1$\sigma$ errors for different cosmological models.}
\label{tab:params}
\end{table}

\fi

\subsection{Goodness of fit}
\label{sec:aic-bic-criteria}

In assessing the goodness of fit with several models, comparing those with different number of fit parameters can pose challenges both qualitatively and quantitatively. One straightforward method involves comparing the differences in the minima of the $\chi^2$ values ($\chi^2_{\rm min}$) between two models, with one serving as the base model, and also computing the $\chi^2_{\rm min}$ per degree of freedom. For the models studied in the present work, the differences $\Delta\chi^2=\chi^2_{\rm ref} - \chi^2_{\rm model}$ relative to the \emph{reference} flat $\Lambda$CDM model are presented in third row from the bottom of Table~\ref{tab:par-val-mcmc} and \ref{tab:par-val-mcmc-lcdm}.

Expanding the parameter space to describe observed data typically results in lower  $\chi^2_{\rm min}$ values. Therefore, a more effective approach to identifying the most suitable model for the observed data should account for the additional parameters introduced during model fitting relative to a base model. One commonly used model selection criterion, derived from information theory, is the Akaike Information Criterion (AIC). Alongside the minimum of $\chi^2=-2\ln(\mathcal{L})$, where $\mathcal{L}$ represents the likelihood function of the model, the AIC formula incorporates an additional penalty term dependent on the number of parameters `$k$' that were used to fit the data. It is given by~\cite{AIC1974},
\begin{equation}
{\rm AIC} = 2k - 2\ln(\mathcal{L_{\rm max}}) = 2k + \chi^2_{\rm min}\,.
\end{equation}
Here, $\chi^2_{\rm min}$ represents the minimum $\chi^2$ value attained in the parameter space after optimization of Eq.~(\ref{eq:chi2-cov}). Relative AIC values for the candidate models are then calculated using one of the models as a reference or base model. This provides an estimate of the probability of the other models in effectively describing the observed data relative to the chosen reference model (taken to be `1' i.e., the most probable model to describe our data). It is expressed as,
\begin{equation}
{\rm AIC}_{\rm rel} = \exp\left(-\Delta/2\right)\,,
\end{equation}
where $\Delta$ in the above expression is the difference between AIC values of a model with respect to a reference model that is given by,
\begin{equation}
    \Delta = {\rm AIC}_{\rm Bianchi-I} - {\rm AIC}_{\Lambda\rm CDM}\,.
\end{equation}

In our analysis, flat $\Lambda$CDM model serves as base model for comparing the anisotropic Bianchi-I model with anisotropic dark energy and the $w$CDM model. The difference in AICs and relative AIC values, $\Delta$ and ${\rm AIC}_{\rm rel}$ respectively, compared to standard concordance model are listed at the end of Table~\ref{tab:par-val-mcmc} and \ref{tab:par-val-mcmc-lcdm}. While the ${\rm AIC}_{\rm rel}$ values indicate how probable various models are under study in comparison to the chosen base model, the difference in AIC values, $\Delta$, also enable us in gauging how well a particular candidate model is supported vis-a-vis the reference model. They are given in Table~\ref{tab:AIC-inference}~\cite{Burnham_04}.

\begin{table}[b]
\centering
\begin{tabular}{|c|c|}
\hline
$\Delta$ & Strength of Evidence \\
\hline
$\leq 2$ & Strong support \\
 4 - 7 & Less support \\
$\geq$ 10 & No support \\
\hline
\end{tabular}
\caption{Strength of evidence in support of a model `$i$' based on the difference in AIC values: $\Delta_i={\rm AIC}_i - {\rm AIC}_{\rm ref}$, where ${\rm AIC}_{\rm ref}$ is the AIC of the \emph{reference} model (or the one with lowest AIC value) from among the candidate models being studied~\cite{Burnham_04}.}
\label{tab:AIC-inference}
\end{table}

Our analysis reveals that the $w_b$CDM model is favored over the other anisotropic cases viz., $w_a$CDM, $\bar{w}-\delta_w$CDM and B-I~$\Lambda$CDM. Its favored by the data better than the standard $\Lambda$CDM or the $w$CDM models based on AIC.

In light of the interpretation provided by AIC for model selection~\cite{AIC1974,Burnham_04}, the flat $\Lambda$CDM model is $\approx$1.09, 0.74, 2.8, 4,17, and 2.08 times more (if $>1$)/less ($<1$) probable than $w_a$CDM, $w_b$CDM, $\bar{w}-\delta_w$CDM, B-I~$\Lambda$CDM and $w$CDM models respectively. So there appears to be a preference for anisotropic expansion of our universe along the $z$-axis of the Bianchi-I model (cosmic preferred axis) with anisotropic dark energy equation of state being different from ``$-1$'' normal to the plane of residual planar symmetry i.e., $w_a=-1$ and $w_b=-0.850\pm0.099$.

\section{Conclusions}
\label{sec:concl}
In this study, assuming a Bianchi-I universe, we aimed to estimate the level of anisotropy in the observed universe by analyzing cosmic shear and eccentricity, constraining the density fractions of isotropic (dark+visible) matter and anisotropic dark energy, and identifying any cosmic preferred axis. The Bianchi-I model considered here is an anisotropic cosmological framework with residual planar symmetry in the spatial metric. We rederived the evolution equations for key cosmological parameters as coupled differential equations and conducted an MCMC likelihood analysis, solving these equations alongside the luminosity distance–redshift relation by defining an appropriate likelihood function. For this analysis, we used the latest Pantheon+ Type Ia Supernova dataset.

The Hubble parameter ($h_0$) obtained from our Bianchi-I model analysis—using three different anisotropic dark energy equation-of-state parameterizations and an anisotropic extension of $\Lambda$CDM (Cases I, II, III, and IV in Sec.~\ref{sec:efe-Dl})—is consistent with the value reported by the Pantheon+ collaboration~\cite{scolnic2022}.

Our results indicate a negative cosmic shear, irrespective of its precise value, and a nonzero present-day eccentricity in Cases I, II, and III. In Case IV, where the Bianchi-I background was modeled with isotropic sources, cosmic shear was found to be low, with only an upper bound obtained on the eccentricity parameter. This suggests that assuming isotropic sources, despite an anisotropic background, may lead to faster isotropization of the universe. Nevertheless, the cosmic shear constraints indicate that the universe may be expanding mildly anisotropically.

Analyzing the evolution of energy density fractions over time—solving the evolution equations (Eq.~\ref{eq:b1evoleq}) using the derived constraints as initial values—we find that the shear energy density parameter dominates at early times, even exceeding the radiation density fraction, which contradicts early-universe observations. This discrepancy highlights the need for much tighter constraints on cosmic shear to ensure consistency with current cosmological models. More stringent constraints can be obtained using early-universe probes such as the CMB and BBN~\cite{saadeh2016MNRAS,Akarsu:2019}.

The anisotropy axes identified in all four cases are consistent with each other and align with previous findings~\cite{anshul2024}, where the same Bianchi-I model was studied with cosmic strings and magnetic fields as sources of anisotropy.

Interestingly, the constraints on the dark energy equation of state ($\bar{w}$ or $w_{\rm de}$) in the three anisotropic Bianchi-I cases, as well as in the $w$CDM model, favor values larger than ``$-1$'', which is in agreement with recent results from the DESI collaboration~\cite{DESI2024eos}. However, since we did not consider a scenario where the dark energy equation of state evolves over time, this coincidence may arise from the additional parameter freedom in the anisotropic cosmological model studied.

Assessing the evidence for different models, we find that the $w_b$CDM model—where the dark energy equation of state varies along the $z$-axis (cosmic preferred axis)—is preferred over both the flat $\Lambda$CDM and the $w$CDM models with isotropic dark energy. The relative AIC values suggest that while other anisotropic models remain viable, the B-I~$\Lambda$CDM model (an anisotropic background with isotropic standard model sources) is the least favored among the models considered.

Our results indicate that a Bianchi-I universe with anisotropic dark energy as the primary source of anisotropy warrants further investigation. Additional observational data, particularly at intermediate and high redshifts, will be crucial for imposing tighter constraints and making definitive conclusions about these alternative anisotropic cosmological models.

\section*{Acknowledgements}
AV acknowledges the financial support received through research fellowship awarded by Council of Scientific \& Industrial Research (CSIR), India during the course of this project.
We acknowledge National Supercomputing Mission (NSM) for providing computing resources of `PARAM
Shivay' at Indian Institute of Technology (BHU), Varanasi, which is implemented by C-DAC and supported by the Ministry of Electronics and Information Technology (MeitY) and Department of Science and Technology (DST), Government of India..
DFM thanks the Research Council of Norway for their support and the resources provided by UNINETT Sigma2 - the National Infrastructure for High Performance Computing and Data Storage in Norway.
Software acknowledgments:
\texttt{Cobaya}\footnote{\url{https://cobaya.readthedocs.io/}}~\cite{torrado2021},
\texttt{GetDist}\footnote{\url{https://cobaya.readthedocs.io/en/latest/}}~\cite{lewis:2019xzd},
\texttt{SciPy}\footnote{\url{https://scipy.org/}}~\cite{scipy2020},
\texttt{NumPy}\footnote{\url{https://numpy.org/}}~\cite{numpy2020},
\texttt{Astropy}\footnote{\url{http://www.astropy.org}}~\cite{astropy2013,astropy2018,astropy2022},
and \texttt{Matplotlib}\footnote{\url{https://matplotlib.org/}}~\cite{hunter:2007}.

\bibliographystyle{unsrt}
\bibliography{ref_ade_etal}

\begin{thebibliography}{10}

\bibitem{WeinbergGRCosmoBook}
Steven Weinberg.
\newblock {\em {Gravitation and Cosmology: Principles and Applications of the
  General Theory of Relativity}}.
\newblock Wiley, New York, NY, 1972.

\bibitem{schwarz2016}
Dominik~J. {Schwarz}, Craig~J. {Copi}, Dragan {Huterer}, and Glenn~D.
  {Starkman}.
\newblock {CMB anomalies after Planck}.
\newblock {\em Classical and Quantum Gravity}, 33(18):184001, September 2016.

\bibitem{bull2016}
Philip Bull et~al.
\newblock {Beyond {\ensuremath{\Lambda}}CDM: Problems, solutions, and the road
  ahead}.
\newblock {\em Phys. Dark Univ.}, 12:56--99, 2016.

\bibitem{perivolaropoulos2022}
L.~{Perivolaropoulos} and F.~{Skara}.
\newblock {Challenges for {\ensuremath{\Lambda}}CDM: An update}.
\newblock {\em \nar}, 95:101659, December 2022.

\bibitem{abdalla2022}
Elcio Abdalla et~al.
\newblock {Cosmology intertwined: A review of the particle physics,
  astrophysics, and cosmology associated with the cosmological tensions and
  anomalies}.
\newblock {\em JHEAp}, 34:49--211, 2022.

\bibitem{aluri2023cp}
Pavan~Kumar Aluri et~al.
\newblock {Is the observable Universe consistent with the cosmological
  principle?}
\newblock {\em Class. Quant. Grav.}, 40(9):094001, 2023.

\bibitem{cobe1996maps}
C.~L. {Bennett}, A.~J. {Banday}, K.~M. {Gorski}, G.~{Hinshaw}, P.~{Jackson},
  P.~{Keegstra}, A.~{Kogut}, G.~F. {Smoot}, D.~T. {Wilkinson}, and E.~L.
  {Wright}.
\newblock {Four-Year COBE DMR Cosmic Microwave Background Observations: Maps
  and Basic Results}.
\newblock {\em \apjl}, 464:L1, June 1996.

\bibitem{wmap9yrmaps}
C.~L. Bennett et~al.
\newblock {Nine-Year Wilkinson Microwave Anisotropy Probe (WMAP) Observations:
  Final Maps and Results}.
\newblock {\em Astrophys. J. Suppl.}, 208:20, 2013.

\bibitem{plk2018maps}
N.~Aghanim et~al.
\newblock {Planck 2018 results. I. Overview and the cosmological legacy of
  Planck}.
\newblock {\em Astron. Astrophys.}, 641:A1, 2020.

\bibitem{wmap7yranom}
C.~L. Bennett et~al.
\newblock {Seven-Year Wilkinson Microwave Anisotropy Probe (WMAP) Observations:
  Are There Cosmic Microwave Background Anomalies?}
\newblock {\em Astrophys. J. Suppl.}, 192:17, 2011.

\bibitem{plk2013isostat}
P.~A.~R. Ade et~al.
\newblock {Planck 2013 results. XXIII. Isotropy and statistics of the CMB}.
\newblock {\em Astron. Astrophys.}, 571:A23, 2014.

\bibitem{plk2015isostat}
P.~A.~R. Ade et~al.
\newblock {Planck 2015 results. XVI. Isotropy and statistics of the CMB}.
\newblock {\em Astron. Astrophys.}, 594:A16, 2016.

\bibitem{plk2018isostat}
Y.~Akrami et~al.
\newblock {Planck 2018 results. VII. Isotropy and Statistics of the CMB}.
\newblock {\em Astron. Astrophys.}, 641:A7, 2020.

\bibitem{schwarz2007sn1a}
D.~J. {Schwarz} and B.~{Weinhorst}.
\newblock {(An)isotropy of the Hubble diagram: comparing hemispheres}.
\newblock {\em \aap}, 474(3):717--729, November 2007.

\bibitem{perivolaropoulos2010sn1a}
I.~{Antoniou} and L.~{Perivolaropoulos}.
\newblock {Searching for a cosmological preferred axis: Union2 data analysis
  and comparison with other probes}.
\newblock {\em \jcap}, 2010(12):012, December 2010.

\bibitem{subir2011sn1a}
Jacques {Colin}, Roya {Mohayaee}, Subir {Sarkar}, and Arman {Shafieloo}.
\newblock {Probing the anisotropic local Universe and beyond with SNe Ia data}.
\newblock {\em \mnras}, 414(1):264--271, June 2011.

\bibitem{Kalus:2013}
B.~{Kalus}, D.~J. {Schwarz}, M.~{Seikel}, and A.~{Wiegand}.
\newblock {Constraints on anisotropic cosmic expansion from supernovae}.
\newblock {\em \aap}, 553:A56, May 2013.

\bibitem{wiltshire2013sn1a}
David~L. {Wiltshire}, Peter~R. {Smale}, Teppo {Mattsson}, and Richard
  {Watkins}.
\newblock {Hubble flow variance and the cosmic rest frame}.
\newblock {\em \prd}, 88(8):083529, October 2013.

\bibitem{2014wang}
J.~S. {Wang} and F.~Y. {Wang}.
\newblock {Probing the anisotropic expansion from supernovae and GRBs in a
  model-independent way}.
\newblock {\em \mnras}, 443(2):1680--1687, September 2014.

\bibitem{appleby2015sn1a}
Stephen {Appleby}, Arman {Shafieloo}, and Andrew {Johnson}.
\newblock {Probing Bulk Flow with Nearby SNe Ia Data}.
\newblock {\em \apj}, 801(2):76, March 2015.

\bibitem{Soltis:2019}
John {Soltis}, Arya {Farahi}, Dragan {Huterer}, and C.~Michael {Liberato}.
\newblock {Percent-Level Test of Isotropic Expansion Using Type Ia Supernovae}.
\newblock {\em \prl}, 122(9):091301, March 2019.

\bibitem{zhao2019sn1a}
Dong {Zhao}, Yong {Zhou}, and Zhe {Chang}.
\newblock {Anisotropy of the Universe via the Pantheon supernovae sample
  revisited}.
\newblock {\em \mnras}, 486(4):5679--5689, July 2019.

\bibitem{hutsemekers1998}
D.~{Hutsemekers}.
\newblock {Evidence for very large-scale coherent orientations of quasar
  polarization vectors}.
\newblock {\em \aap}, 332:410--428, April 1998.

\bibitem{hutsemekers2001}
D.~{Hutsem{\'e}kers} and H.~{Lamy}.
\newblock {Confirmation of the existence of coherent orientations of quasar
  polarization vectors on cosmological scales}.
\newblock {\em \aap}, 367:381--387, February 2001.

\bibitem{jain2004}
Pankaj {Jain}, Gaurav {Narain}, and S.~{Sarala}.
\newblock {Large-scale alignment of optical polarizations from distant QSOs
  using coordinate-invariant statistics}.
\newblock {\em \mnras}, 347(2):394--402, January 2004.

\bibitem{hutsemekers2014}
D.~{Hutsem{\'e}kers}, L.~{Braibant}, V.~{Pelgrims}, and D.~{Sluse}.
\newblock {Alignment of quasar polarizations with large-scale structures}.
\newblock {\em \aap}, 572:A18, December 2014.

\bibitem{Secrest2021}
Nathan~J. {Secrest}, Sebastian {von Hausegger}, Mohamed {Rameez}, Roya
  {Mohayaee}, Subir {Sarkar}, and Jacques {Colin}.
\newblock {A Test of the Cosmological Principle with Quasars}.
\newblock {\em \apjl}, 908(2):L51, February 2021.

\bibitem{birch1982}
P.~{Birch}.
\newblock {Is the Universe rotating?}
\newblock {\em \nat}, 298(5873):451--454, July 1982.

\bibitem{jain1999}
Pankaj {Jain} and John~P. {Ralston}.
\newblock {Anisotropy in the Propagation of Radio Polarizations from
  Cosmologically Distant Galaxies}.
\newblock {\em Modern Physics Letters A}, 14(6):417--432, January 1999.

\bibitem{tiwari2013radioalgn}
Prabhakar {Tiwari} and Pankaj {Jain}.
\newblock {Polarization Alignment in Jvas/class Flat Spectrum Radio Surveys}.
\newblock {\em International Journal of Modern Physics D}, 22(12):1350089,
  October 2013.

\bibitem{watkins2009}
Richard {Watkins}, Hume~A. {Feldman}, and Michael~J. {Hudson}.
\newblock {Consistently large cosmic flows on scales of 100h$^{-1}$Mpc: a
  challenge for the standard {\ensuremath{\Lambda}}CDM cosmology}.
\newblock {\em \mnras}, 392(2):743--756, January 2009.

\bibitem{Hoffman2024MNRAS}
Yehuda {Hoffman}, Aurelien {Valade}, Noam~I. {Libeskind}, Jenny~G. {Sorce},
  R.~Brent {Tully}, Simon {Pfeifer}, Stefan {Gottl{\"o}ber}, and Daniel
  {Pomar{\`e}de}.
\newblock {The large-scale velocity field from the Cosmicflows-4 data}.
\newblock {\em \mnras}, 527(2):3788--3805, January 2024.

\bibitem{johnjain2004}
John~P. {Ralston} and Pankaj {Jain}.
\newblock {The Virgo Alignment Puzzle in Propagation of Radiation on
  Cosmological Scales}.
\newblock {\em International Journal of Modern Physics D}, 13(9):1857--1877,
  January 2004.

\bibitem{Perlmutter:1999}
S.~Perlmutter et~al.
\newblock {Measurements of {\ensuremath{\Omega}} and {\ensuremath{\Lambda}}
  from 42 High-Redshift Supernovae}.
\newblock {\em Astrophys. J.}, 517:565--586, 1999.

\bibitem{Spergel2003ApJS}
D.~N. Spergel et~al.
\newblock {First year Wilkinson Microwave Anisotropy Probe (WMAP) observations:
  Determination of cosmological parameters}.
\newblock {\em Astrophys. J. Suppl.}, 148:175--194, 2003.

\bibitem{plk2018cosmopar}
N.~Aghanim et~al.
\newblock {Planck 2018 results. VI. Cosmological parameters}.
\newblock {\em Astron. Astrophys.}, 641:A6, 2020.
\newblock [Erratum: Astron.Astrophys. 652, C4 (2021)].

\bibitem{Armendariz-Picon2004}
C.~{Armend{\'a}riz-Pic{\'o}n}.
\newblock {Could dark energy be vector-like?}
\newblock {\em \jcap}, 2004(7):007, July 2004.

\bibitem{Bohmer2007}
C.~G. {B{\"o}hmer} and T.~{Harko}.
\newblock {Dark energy as a massive vector field}.
\newblock {\em European Physical Journal C}, 50(2):423--429, April 2007.

\bibitem{Rodrigues2008}
Davi~C. {Rodrigues}.
\newblock {Anisotropic cosmological constant and the CMB quadrupole anomaly}.
\newblock {\em \prd}, 77(2):023534, January 2008.

\bibitem{Koivisto2008a}
Tomi {Koivisto} and David~F. {Mota}.
\newblock {Vector field models of inflation and dark energy}.
\newblock {\em \jcap}, 2008(8):021, August 2008.

\bibitem{Koivisto2009}
Tomi~S. {Koivisto} and Nelson~J. {Nunes}.
\newblock {Inflation and dark energy from three-forms}.
\newblock {\em \prd}, 80(10):103509, November 2009.

\bibitem{Appleby2010}
Stephen {Appleby}, Richard {Battye}, and Adam {Moss}.
\newblock {Constraints on the anisotropy of dark energy}.
\newblock {\em \prd}, 81(8):081301, April 2010.

\bibitem{Zuntz2010}
J.~{Zuntz}, T.~G. {Zlosnik}, F.~{Bourliot}, P.~G. {Ferreira}, and G.~D.
  {Starkman}.
\newblock {Vector field models of modified gravity and the dark sector}.
\newblock {\em \prd}, 81(10):104015, May 2010.

\bibitem{Campanelli2011ade}
L.~{Campanelli}, P.~{Cea}, G.~L. {Fogli}, and L.~{Tedesco}.
\newblock {Anisotropic Dark Energy and Ellipsoidal Universe}.
\newblock {\em International Journal of Modern Physics D}, 20(6):1153--1166,
  January 2011.

\bibitem{Thorsrud2012}
Mikjel {Thorsrud}, David~F. {Mota}, and Sigbj{\o}rn {Hervik}.
\newblock {Cosmology of a scalar field coupled to matter and an
  isotropy-violating Maxwell field}.
\newblock {\em Journal of High Energy Physics}, 2012:66, October 2012.

\bibitem{Akarsu2014}
{\"O}zg{\"u}r {Akarsu}, Tekin {Dereli}, and Neslihan {Oflaz}.
\newblock {Accelerating anisotropic cosmologies in Brans-Dicke gravity coupled
  to a mass-varying vector field}.
\newblock {\em Classical and Quantum Gravity}, 31(4):045020, February 2014.

\bibitem{Saha2014}
Bijan Saha.
\newblock {Isotropic and anisotropic dark energy models}.
\newblock {\em Phys. Part. Nucl.}, 45:349--396, 2014.

\bibitem{Lavinia2016}
Lavinia {Heisenberg}, Ryotaro {Kase}, and Shinji {Tsujikawa}.
\newblock {Anisotropic cosmological solutions in massive vector theories}.
\newblock {\em \jcap}, 2016(11):008, November 2016.

\bibitem{Beltran2019}
Juan~P. {Beltr{\'a}n Almeida}, Alejandro {Guarnizo}, Ryotaro {Kase}, Shinji
  {Tsujikawa}, and C{\'e}sar~A. {Valenzuela-Toledo}.
\newblock {Anisotropic 2-form dark energy}.
\newblock {\em Physics Letters B}, 793:396--404, June 2019.

\bibitem{Motoa-Manzano2021}
Josu{\'e} {Motoa-Manzano}, J.~Bayron {Orjuela-Quintana}, Thiago~S. {Pereira},
  and C{\'e}sar~A. {Valenzuela-Toledo}.
\newblock {Anisotropic solid dark energy}.
\newblock {\em Physics of the Dark Universe}, 32:100806, May 2021.

\bibitem{Orjuela-Quintan2021}
J.~Bayron {Orjuela-Quintana} and C{\'e}sar~A. {Valenzuela-Toledo}.
\newblock {Anisotropic k-essence}.
\newblock {\em Physics of the Dark Universe}, 33:100857, September 2021.

\bibitem{Jaffe2005ApJ}
T.~R. {Jaffe}, A.~J. {Banday}, H.~K. {Eriksen}, K.~M. {G{\'o}rski}, and F.~K.
  {Hansen}.
\newblock {Evidence of Vorticity and Shear at Large Angular Scales in the WMAP
  Data: A Violation of Cosmological Isotropy?}
\newblock {\em \apjl}, 629(1):L1--L4, August 2005.

\bibitem{Pontzen2009}
Andrew {Pontzen}.
\newblock {Rogues' gallery: The full freedom of the Bianchi CMB anomalies}.
\newblock {\em \prd}, 79(10):103518, May 2009.

\bibitem{coles2011}
Rockhee {Sung} and Peter {Coles}.
\newblock {Temperature and polarization patterns in anisotropic cosmologies}.
\newblock {\em \jcap}, 2011(6):036, June 2011.

\bibitem{saadeh2016MNRAS}
Daniela {Saadeh}, Stephen~M. {Feeney}, Andrew {Pontzen}, Hiranya~V. {Peiris},
  and Jason~D. {McEwen}.
\newblock {A framework for testing isotropy with the cosmic microwave
  background}.
\newblock {\em \mnras}, 462(2):1802--1811, October 2016.

\bibitem{koivisto2008}
Tomi {Koivisto} and David~F. {Mota}.
\newblock {Anisotropic dark energy: dynamics of the background and
  perturbations}.
\newblock {\em \jcap}, 2008(6):018, June 2008.

\bibitem{2011campanelli}
L.~{Campanelli}, P.~{Cea}, G.~L. {Fogli}, and A.~{Marrone}.
\newblock {Testing the isotropy of the Universe with type Ia supernovae}.
\newblock {\em \prd}, 83(10):103503, May 2011.

\bibitem{2013appleby}
Stephen~A. {Appleby} and Eric~V. {Linder}.
\newblock {Probing dark energy anisotropy}.
\newblock {\em \prd}, 87(2):023532, January 2013.

\bibitem{Wang:2018}
Yu-Yang {Wang} and F.~Y. {Wang}.
\newblock {Testing the isotropy of the Universe with Type Ia supernovae in a
  model-independent way}.
\newblock {\em \mnras}, 474(3):3516--3522, March 2018.

\bibitem{Tedesco2018}
Luigi {Tedesco}.
\newblock {Ellipsoidal expansion of the Universe, cosmic shear, acceleration
  and jerk parameter}.
\newblock {\em European Physical Journal Plus}, 133(5):188, May 2018.

\bibitem{Akarsu:2019}
{\"O}zg{\"u}r {Akarsu}, Suresh {Kumar}, Shivani {Sharma}, and Luigi {Tedesco}.
\newblock {Constraints on a Bianchi type I spacetime extension of the standard
  {\ensuremath{\Lambda}} CDM model}.
\newblock {\em \prd}, 100(2):023532, July 2019.

\bibitem{Amirhashchi:2020}
Hassan Amirhashchi and Soroush Amirhashchi.
\newblock {Constraining Bianchi Type I Universe With Type Ia Supernova and H(z)
  Data}.
\newblock {\em Phys. Dark Univ.}, 29:100557, 2020.

\bibitem{2021yadav}
Anil~Kumar {Yadav}, Avinash~K. {Yadav}, Manvinder {Singh}, Rajendra {Prasad},
  Nafis {Ahmad}, and Kangujam~Priyokumar {Singh}.
\newblock {Constraining a bulk viscous Bianchi type I dark energy dominated
  universe with recent observational data}.
\newblock {\em \prd}, 104(6):064044, September 2021.

\bibitem{2022rahman}
W.~{Rahman}, R.~{Trotta}, S.~S. {Boruah}, M.~J. {Hudson}, and D.~A. {van Dyk}.
\newblock {New constraints on anisotropic expansion from supernovae Type Ia}.
\newblock {\em \mnras}, 514(1):139--163, July 2022.

\bibitem{2023Akarsu}
{\"O}zg{\"u}r {Akarsu}, Eleonora {Di Valentino}, Suresh {Kumar}, Maya
  {{\"O}zyi{\u{g}}it}, and Shivani {Sharma}.
\newblock {Testing spatial curvature and anisotropic expansion on top of the
  {\ensuremath{\Lambda}}CDM model}.
\newblock {\em Physics of the Dark Universe}, 39:101162, February 2023.

\bibitem{maccallumellis1969}
G.~F.~R. Ellis and Malcolm A.~H. MacCallum.
\newblock {A Class of homogeneous cosmological models}.
\newblock {\em Commun. Math. Phys.}, 12:108--141, 1969.

\bibitem{maccallumellis1970}
M.~A.~H. {MacCallum} and G.~F.~R. {Ellis}.
\newblock {A class of homogeneous cosmological models: II. Observations}.
\newblock {\em Communications in Mathematical Physics}, 19(1):31--64, March
  1970.

\bibitem{RyanShepley1975}
Michael~P. {Ryan} and Lawrence~C. {Shepley}.
\newblock {\em {Homogeneous Relativistic Cosmologies}}.
\newblock Princeton Series in Physics. Princeton University Press, Princeton,
  1975.

\bibitem{Wainwright1997}
J.~{Wainwright} and G.~F.~R. {Ellis}.
\newblock {\em {Dynamical Systems in Cosmology}}.
\newblock Cambridge Univ. Press, Cambridge, 1997.

\bibitem{Stephani2003}
Hans {Stephani}, Dietrich {Kramer}, Malcolm {MacCallum}, Cornelius
  {Hoenselaers}, and Eduard {Herlt}.
\newblock {\em {Exact solutions of Einstein's field equations}}.
\newblock Cambridge Monographs on Mathematical Physics. Cambridge Univ. Press,
  Cambridge, 2003.

\bibitem{Coley2003}
A.~A. {Coley}.
\newblock {\em {Dynamical systems and cosmology}}.
\newblock Astrophysics and Space Science Library. Kluwer Academic Publishers,
  Dordrecht, Netherlands, 2003.

\bibitem{barrow1997}
John~D. {Barrow}.
\newblock {Cosmological limits on slightly skew stresses}.
\newblock {\em \prd}, 55(12):7451--7460, June 1997.

\bibitem{barera2004}
Arjun {Berera}, Roman~V. {Buniy}, and Thomas~W. {Kephart}.
\newblock {The eccentric universe}.
\newblock {\em \jcap}, 2004(10):016, October 2004.

\bibitem{campanelli2009}
Leonardo {Campanelli}.
\newblock {Model of universe anisotropization}.
\newblock {\em \prd}, 80(6):063006, September 2009.

\bibitem{campanelli2006}
L.~{Campanelli}, P.~{Cea}, and L.~{Tedesco}.
\newblock {Ellipsoidal Universe Can Solve the Cosmic Microwave Background
  Quadrupole Problem}.
\newblock {\em \prl}, 97(13):131302, September 2006.

\bibitem{aluri2013sn1a}
Pavan~K. {Aluri}, Sukanta {Panda}, Manabendra {Sharma}, and Snigdha {Thakur}.
\newblock {Anisotropic universe with anisotropic sources}.
\newblock {\em \jcap}, 2013(12):003, December 2013.

\bibitem{anshul2024}
Anshul {Verma}, Sanjeet~K. {Patel}, Pavan~K. {Aluri}, Sukanta {Panda}, and
  David~F. {Mota}.
\newblock {Constraints on Bianchi-I type universe with SH0ES anchored Pantheon+
  SNIa data}.
\newblock {\em \jcap}, 2024(6):071, June 2024.

\bibitem{davis2011}
Tamara~M. Davis et~al.
\newblock {The Effect of Peculiar Velocities on Supernova Cosmology}.
\newblock {\em Astrophys. J.}, 741:67, 2011.

\bibitem{scolnic2022}
Dan Scolnic et~al.
\newblock {The Pantheon+ Analysis: The Full Data Set and Light-curve Release}.
\newblock {\em Astrophys. J.}, 938(2):113, 2022.

\bibitem{riess2022}
Adam~G. Riess et~al.
\newblock {A Comprehensive Measurement of the Local Value of the Hubble
  Constant with 1 km s$^{-1}$ Mpc$^{-1}$ Uncertainty from the Hubble Space
  Telescope and the SH0ES Team}.
\newblock {\em Astrophys. J. Lett.}, 934(1):L7, 2022.

\bibitem{Barrow:1976}
John {Barrow}.
\newblock {Light elements and the isotropy of the Universe}.
\newblock {\em \mnras}, 175:359--370, May 1976.

\bibitem{Campanelli2011He4}
Leonardo {Campanelli}.
\newblock {Helium-4 synthesis in an anisotropic universe}.
\newblock {\em \prd}, 84(12):123521, December 2011.

\bibitem{AIC1974}
H.~{Akaike}.
\newblock {A New Look at the Statistical Model Identification}.
\newblock {\em IEEE Transactions on Automatic Control}, 19:716--723, January
  1974.

\bibitem{Burnham_04}
K.~P. Burnham and D.~R. Anderson.
\newblock Multimodel inference: {Understanding} {AIC} and {BIC} in model
  selection.
\newblock {\em Sociological Methods and Research}, 33(2):261--304, November
  2004.

\bibitem{akrami2014}
Y.~{Akrami}, Y.~{Fantaye}, A.~{Shafieloo}, H.~K. {Eriksen}, F.~K. {Hansen},
  A.~J. {Banday}, and K.~M. {G{\'o}rski}.
\newblock {Power Asymmetry in WMAP and Planck Temperature Sky Maps as Measured
  by a Local Variance Estimator}.
\newblock {\em \apjl}, 784(2):L42, April 2014.

\bibitem{bengaly2018}
Carlos A.~P. {Bengaly}, Roy {Maartens}, and Mario~G. {Santos}.
\newblock {Probing the Cosmological Principle in the counts of radio galaxies
  at different frequencies}.
\newblock {\em \jcap}, 2018(4):031, April 2018.

\bibitem{alpha_dipole}
Antonio {Mariano} and Leandros {Perivolaropoulos}.
\newblock {Is there correlation between fine structure and dark energy cosmic
  dipoles?}
\newblock {\em \prd}, 86(8):083517, October 2012.

\bibitem{DESI2024eos}
A.~G. Adame et~al.
\newblock {DESI 2024 VI: Cosmological Constraints from the Measurements of
  Baryon Acoustic Oscillations}.
\newblock 4 2024.

\bibitem{torrado2021}
Jes{\'u}s {Torrado} and Antony {Lewis}.
\newblock {Cobaya: code for Bayesian analysis of hierarchical physical models}.
\newblock {\em \jcap}, 2021(5):057, May 2021.

\bibitem{lewis:2019xzd}
Antony Lewis.
\newblock {GetDist: a Python package for analysing Monte Carlo samples}.
\newblock 2019.

\bibitem{scipy2020}
Pauli Virtanen et~al.
\newblock {{SciPy} 1.0: Fundamental Algorithms for Scientific Computing in
  Python}.
\newblock {\em Nature Methods}, 17:261--272, 2020.

\bibitem{numpy2020}
Charles~R. Harris et~al.
\newblock Array programming with {NumPy}.
\newblock {\em Nature}, 585(7825):357--362, September 2020.

\bibitem{astropy2013}
T.~P. {Robitaille} et~al.
\newblock {Astropy: A community Python package for astronomy}.
\newblock {\em \aap}, 558:A33, October 2013.

\bibitem{astropy2018}
A.~M. {Price-Whelan} et~al.
\newblock {The Astropy Project: Building an Open-science Project and Status of
  the v2.0 Core Package}.
\newblock {\em \aj}, 156(3):123, September 2018.

\bibitem{astropy2022}
Adrian~M. {Price-Whelan} et~al.
\newblock {The Astropy Project: Sustaining and Growing a Community-oriented
  Open-source Project and the Latest Major Release (v5.0) of the Core Package}.
\newblock {\em apj}, 935(2):167, August 2022.

\bibitem{hunter:2007}
J.~D. Hunter.
\newblock Matplotlib: A 2d graphics environment.
\newblock {\em Computing in Science \& Engineering}, 9(3):90--95, 2007.

\end{thebibliography}

\end{document}